\documentclass[
superscriptaddress,
amsmath,
amssymb,
achemso,
aps,
prl,
twocolumn,
]{revtex4-2}

\usepackage{braket}
\usepackage{graphicx}
\usepackage[colorlinks=true, allcolors=blue]{hyperref}
\usepackage{gensymb}
\usepackage{derivative}
\usepackage{upgreek}
\usepackage{float}  
\usepackage{xcolor}
\usepackage{footmisc}
\usepackage{siunitx}
\usepackage{calrsfs}
\usepackage{mathrsfs}
\usepackage{footmisc}

\usepackage[colorlinks=true, allcolors=blue]{hyperref}
\usepackage{mathtools}
\usepackage[normalem]{ulem}

\newcommand{\ca}{$^{40}$Ca$^+$}

\newsavebox\mcFcontent
\savebox\mcFcontent{$\mathcal{F}$}

\DeclarePairedDelimiter{\evdel}{\langle}{\rangle}
\newcommand{\ev}{\operatorname{}\evdel}

\begin{document}
\title{Nonlinear-enhanced wideband sensing via subharmonic excitation of a quantum harmonic oscillator}

\author{Hao Wu}
\email{hao.wu@physics.ucla.edu}
\thanks{These authors contributed equally to this work}
\author{Clayton Z. C. Ho}
\thanks{These authors contributed equally to this work}
\author{Grant D. Mitts}
\thanks{These authors contributed equally to this work}
\author{Joshua A. Rabinowitz}
\author{Eric R. Hudson}

\affiliation{Department of Physics and Astronomy, University of California Los Angeles, Los Angeles, CA, USA}
\affiliation{Challenge Institute for Quantum Computation, University of California Los Angeles, Los Angeles, CA, USA}
\affiliation{Center for Quantum Science and Engineering, University of California Los Angeles, Los Angeles, CA, USA}
    
\date{\today} 

\begin{abstract}
A key advantage of quantum metrology is the ability to surpass the standard quantum limit~(SQL) for measurement precision through the use of non-classical states.
However, there is typically little to no improvement in precision with the use of non-classical states for measurements whose duration exceeds the decoherence time of the underlying quantum states. 
Measurements aimed at the ultimate possible precision are thus performed almost exclusively with classical states and, therefore, are constrained by the SQL.
Here, we demonstrate that by using the phenomenon of subharmonic excitation, in combination with a recently
demonstrated technique of Raman excitation of a harmonic oscillator, the frequency of an electric field can be measured at a resolution below the SQL of the corresponding linear generator. With this method we measure a radio-frequency electrical signal with a fractional frequency uncertainty of 0.56~Hz/80~MHz=7e-9 , which to our knowledge is the most precise frequency measurement of a radio-frequency electrical signal using a quantum harmonic oscillator.
Because the input states can be classical, the coherence time is not degraded by the enhanced decoherence typically associated with nonclassical states, thereby improving the ultimate achievable precision.
While we demonstrate this technique using motional Raman subharmonic excitation of a single \ca\ ion through engineered Floquet states, this technique is expected to be extendable to other platforms, such as NV centers, solid-state qubits, and neutral atoms, where it can provide metrological gain for sensing across the radio frequency, microwave, and optical domains. 
\end{abstract}

\maketitle


Quantum harmonic oscillators (QHOs) are highly-sensitive platforms for detecting optical and radio-frequency (RF) electromagnetic fields near the natural frequency $\omega$ of the oscillator~\cite{Hempel2013, Burd2019, Wolf2019}.
QHOs are thus powerful precision sensors in areas as diverse as ultra-precise measurements of force~\cite{Knunz2010, Biercuk2010, Gilmore2017, Liu2021} and acceleration~\cite{Campbell2017}, searches of axionic dark matter~\cite{Bradley2003}, and gravitational waves detection~\cite{LIGO}. 
Importantly, the recent development of motional Raman protocols have unlocked the potential of QHOs for broadband field sensing by extending the usable frequency range by over three orders of magnitude across the RF and microwave bands~\cite{Wu2025}.

The sensitivity of a QHO-based measurement to a physical parameter $\theta$ is bounded by $\sigma_\theta = (\sqrt{F_q(\phi)} (\partial \phi/\partial \theta))^{-1}$, where $F_q{\left(\phi\right)}$ is the Fisher information for parameter $\phi$ of a QHO in state $q$.
For coherent states $\ket{\alpha}$, the displacement $\alpha$ is typically a product of the interrogation time $\tau$ and some function $f$ that depends on the frequency $\omega_{s}$ and strength $\Omega$ of an applied field (referred to hereafter as the signal field), i.e. $\alpha = \tau f{\left(\omega_{s},\Omega\right)}$.
Thus, given that the Fisher information for a protocol that measures QHO displacement ($\alpha$) $F_\alpha \leq 4$, the standard quantum limit (SQL) bounds the sensitivity of measurements of the signal frequency to $\sigma_{\omega_s} = (2\tau (\partial f/\partial \omega_s))^{-1}$.
Further limitations appear in systems with finite coherence time $\tau_c$ where sensitivity is optimized at $\tau \approx \tau_c/2$~\cite{Huelga1997}.

To improve sensitivity, non-classical states are employed, such as Schrödinger cat states~\cite{Hempel2013, Johnson2017, Monroe1996}, squeezed states~\cite{Burd2019, Wu2025}, and high Fock states~\cite{Wolf2019, McCormick2019}, to increase the Fisher information~\cite{Wolfthesis}.  
However, these states are challenging to generate and control, as their coherence times scale inversely with metrological gain~\cite{Huelga1997}. In practice, extracting gains $>10 \textrm{ dB}$ with non-classical states, for even short timescales, has proven difficult~\cite{Pezze2018}.

Here, we instead leverage a nonlinear measurement as an alternative means to enhance sensitivity.
It has been theoretically proposed that, by
exploiting $k$-body (nonlinear) probe dynamics, the precision of parameter estimation can surpass the conventional 1/$N$  “Heisenberg” scaling associated with linear (one-body) metrology, instead following a scale of $1/N^{k}$ with optimal entangled inputs, and $1/N^{k-1/2}$ scaling with product inputs~\cite{Boixo2007,Boixo2008,Boixo2009,Huang2024}.
In contrast, we instead propose and experimentally demonstrate an alternative nonlinear approach.
We show that the precision achieved with our nonlinear generator surpasses the SQL of the corresponding linear generator.
Specifically, by parametrically exciting a QHO subharmonic resonance at $2\omega/K$, where $\omega$ is the QHO resonance and $K \in \mathbb{N} $, using a motional Raman protocol~\cite{Wu2025}, we demonstrate that the frequency precision of a signal field can be improved by $\sigma_{\omega_s} = (K \tau (\partial f/\partial \omega_s))^{-1}$.
This $K/2$ sensitivity enhancement arises from a $K/2$-photon process and requires only classical states, entailing no cost to the coherence time $\tau$ regardless of the order $K$.
To be clear, our protocol employs classical probe states and therefore does not surpass the SQL of the nonlinear generator, which improves the scaling by a factor of K/2 relative to the SQL of corresponding linear generator, like K=2 case \cite{SI}.
\nocite{JOHANSSON2012,JOHANSSON2013,Lee2005,Zurek2001,Degen2017,McCormickDisplacedFock2019,Burd2021}

We demonstrate this technique on the motional modes of a single, trapped $^{40}\textrm{Ca}^+$ ion and confirm that subharmonic excitation technique is a $K$-order nonlinear effect via motional Ramsey spectroscopy.
We showcase wideband operation by sensing fields at various signal frequencies, observing the expected $2/K$ sensitivity scaling up to $K = 24$.
Finally, we benchmark the ultimate sensitivities by conducting frequency Allan deviations at various $K$, finding improvement of $12.3(9) \textrm{ dB}$ over the SQL of K=2; in all cases, the best sensitivity achievable with the corresponding linear generator
using classical state is surpassed without use of non-classical states.
We achieve a minimum frequency uncertainty of $0.56(32)\textrm{ Hz}$, representing a sixfold improvement over previous measurements~\cite{Wu2025} to set a new bar for the state-of-the-art in QHO-based frequency sensing in the RF regime.

\begin{figure*}
    \centering
    \includegraphics[width = 1\textwidth]{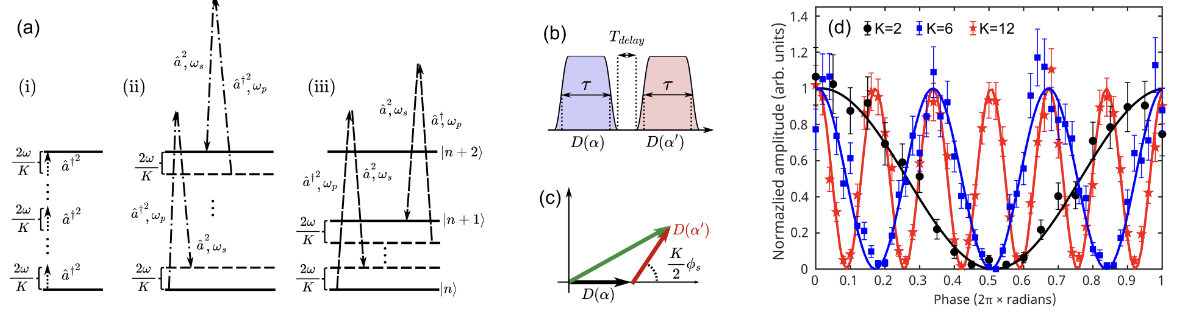}
    \caption{
    Energy-level diagrams for motional subharmonic excitation and verification via motional Ramsey spectroscopy.
    (a) Level diagrams for a $K{\textrm{th}}$-order parametric excitation (i), quadrupole subharmonic Raman excitation (ii) and quadrupole/dipole subharmonic Raman excitation (iii).
    (b-d) Experimental verification of accelerated dynamics via a motional Ramsey sequence with a higher-order subharmonic excitation.
    (b) Simplified pulse sequence of the motional Ramsey scheme used to examine the phase response.
    A subharmonic excitation pulse is applied for interrogation time $\tau$ to generate a displacement $\hat{D}{\left( \alpha \right)}$.
    Following free evolution for time $T_{\textrm{delay}}$, a second subharmonic excitation pulse of equal area is applied with phase $\phi_{s}$ relative to the initial pulse to generate a second displacement $\hat{D}{\left( \alpha'\right)}$, mapping the phase onto the ion's motional state.
    (c) Motional Ramsey protocol in phase space. The phase $\phi_{s}$ rotates the second displacement $\hat{D}{\left( \alpha' \right)}$ relative to the first. The phase dependence at order $K$ thus corresponds to the periodicity of the Ramsey fringes.
    (d) Ramsey fringes generated by scanning the phase $\phi_{s}$ of the second subharmonic excitation pulse at different orders $K$.
    Here, $\tau = 1 \textrm{ ms}$ with delay time is $T_{\textrm{delay}}$=$\SI{50}{\micro s}$.
    Solid lines are fits of the data to a sine.
    Excitation amplitudes for all subharmonic orders $K$ are normalized for comparison.
    Error bars represent one standard error.
    }
    \label{fig:level_diagram}
\end{figure*}

Subharmonic excitation of a harmonic oscillator is driven by a force $\left|\vec{F}\right| = x \left(\Omega/2\right) \cos{\left(\omega_p t\right)}$, where the amplitude is proportional to the equilibrium displacement $x$.
The resultant potential $V = x^{2} \ \Omega\cos{\left(\omega_p t\right)}$ modulates the oscillator's natural frequency $\omega$ and excites a parametric resonance when $\omega_p = 2\omega/K$ for $K \in \mathbb{N}$~\cite{Landau1976Mechanics}.
This process, equivalent to $K$-quanta absorption from the drive field to produce two phonons~(see Fig.~\ref{fig:level_diagram}~(a)), results in a time evolution approximated by a QHO squeezing operator $\hat{S}{\left(r\right)}$, where $r \propto \Omega^{K}/(\omega / 2)^{K-1}$~\cite{SI}.
Subharmonics have been explored in mass spectrometry, where oscillating electric-field gradients were used to excite subharmonics and infer ion mass~\cite{SUDAKOV2000, Zhao2002, COLLINGS2000, Tommaseo2003}.
Resonance occurs at $2\omega = K (\omega_p^{(0)} + \delta)$, with $\omega_p^{(0)} = 2\omega/K$ and detuning $\delta$.
Since linewidth of a parametric resonance is a function of the drive frequency, it also becomes reduced by $K$.
Though these past studies have observed linewidths below the Fourier transform limit (FTL), no improvement in mass resolution was observed.
This is because, though the central frequency $\omega_p^{\left(0\right)}$ is determined with $K$-fold improved sensitivity, i.e. $\partial f/\partial \omega_p  \rightarrow K\partial f/\partial \omega_p$, the parametric excitation is no more sensitive to $\omega$, since $\sqrt{F{\left(\omega_p\right)}}= \sqrt{F{\left(\omega\right)}}/K$.
Subharmonic excitation and its concomitant line-narrowing has also been observed with multi-photon Raman transitions in ultracold atoms~\cite{Trebino1987, Cataliotti2001}, albeit with a different underlying mechanism; as in the mass spectroscopy case, no metrological gain was achieved.

To overcome this limitation of subharmonics, we employ motional Raman transitions to heterodyne the signal tone with an probe tone, exciting a subharmonic in the QHO baseband.
Given the ability to choose the applied probe tone frequency $\omega_p$, this technique allows broadband sensing of $\omega_s$ with sensitivity $K/2$ below the SQL of corresponding linear case (K=2).

Consider the following situation.
A ``signal" tone with voltage $V_{s}$ and phase $\phi_{s}$ at frequency $\omega_{s}$ in quadrupolar configuration (i.e. a linear field gradient) is applied to the ion.
A ``probe" tone at frequency $\omega_p=\omega_s+2\omega_{r/z}/K+\delta$ is then applied with a dipole component (i.e. a uniform field) $V_{d}$ and quadrupole component $V_{ q}$, where $\omega_{r/z}$ is the QHO resonance frequency on mode $r$, $z$.
The Hamiltonian of the ion motion under these fields is:
\begin{align}
    \frac{\hat{H}}{\hbar}&= \omega_{r/z} \hat{a}^{\dagger} \hat{a} + \Omega_{s} \left( \hat{a} + \hat{a}^{\dagger} \right)^2
    \cos{ \left( \omega_{s} t + \phi_{s} \right)}
     \notag \\
    &+ \Omega_{d} \left( \hat{a}+\hat{a}^\dagger \right) \cos{ \left(  \omega_p t \right)}
    + \Omega_{q} \left( \hat{a} + \hat{a}^{\dagger} \right)^2
    \cos{ \left(  \omega_p t \right)},
    \label{eq:Hamil_base}
\end{align}
where $\Omega_{q} = e V_{q} \eta^{2}/\hbar$, $\Omega_{d} =  eV_{d} \eta/\hbar$, $\Omega_{s} = eV_{s}\eta^{2}/\hbar$, and $\eta$ is the effective Lamb-Dicke parameter.
The dipole tone $\Omega_d$ is included as it carries a lower order of $\eta$ than quadrupole tones and modifies the motional Raman interaction to produce a displacement, rather than a squeeze operator.
This is beneficial as coherent states are more easily manipulated and, given our readout procedure, provide significantly higher Fisher information than squeezed states~\cite{SI}.

In the interaction picture with respect to the bare QHO, the Hamiltonian becomes a sum of eight Hermitian components, each comprised of a factor $\Omega_i$, ladder operators, and time dependent phase $e^{\pm i \omega_i t}$, where
$\{\Omega_1 = \Omega_q, \omega_1=\omega_p+2\omega_{r/z}\}$,
$\{\Omega_2 = \Omega_q, \omega_2=\omega_p-2\omega_{r/z}\}$,
$\{\Omega_3 = \Omega_q, \omega_3=\omega_p\}$,
$\{\Omega_4 = \Omega_d, \omega_4=\omega_p+\omega_{r/z}\}$,
$\{\Omega_5 = \Omega_d, \omega_5=\omega_p-\omega_{r/z}\}$,
$\{\Omega_6 = \Omega_s, \omega_6=\omega_s+2\omega_{r/z}\}$,
$\{\Omega_7 = \Omega_s, \omega_7=\omega_s-2\omega_{r/z}\}$,
and $\{\Omega_8 = \Omega_s, \omega_8=\omega_s\}$~\cite{SI}.

The resultant dynamics can be analyzed via multimode Floquet theory~\cite{Ernst2005,Wang2022}, where the process can be understood as driving virtual motional transitions in a multimode Floquet space.
The resulting effective Hamiltonian can be represented as an eight-mode Floquet Hamiltonian~\cite{Ernst2005,Wang2022}:
\begin{align}
    \hat{H} = \sum_{K=1}^\infty \sum_{i=1}^8 & \left(\hat{H}_{1}^{(i)} + \hat{H}_2^{(i)} + \dots + \hat{H}_K^{(i)} \dots \right)\nonumber\\
    &\times e^{-\imath (\sum_{k=1}^8 c_k \omega_k)t}e^{-\imath (c_6+c_7+c_8)\phi_s}
    \label{eq:floquet}
\end{align}

While lower-order terms are readily obtained using methods from Refs.~\cite{Ernst2005,Wang2022}, deriving general analytical expressions for the $K$th-order term $\hat{H}_K^{(i)}$ is nontrivial.
Nonetheless, the essential behaviors are evident from Eqn.~\eqref{eq:floquet}.
A resonant displacement results when the $\sum_k c_k \omega_k = 0$.
Since $c_k \in \mathbb{Z}$, this results in the following restrictions: for the probe tones, $\sum_{k=1}^5 c_k= \pm K/2$, and for the signal tone, $\sum_{k=6}^8 c_k= \mp K/2.$

This result has several consequences.
First, $\delta$ becomes amplified by a factor $\sum_{k=1}^5 c_k = K/2$, narrowing the linewidth relative to a linear measurement by $2/K$.
Second, the $K$th-order Hamiltonian $H_{K}^{(i)}$ involves $K-1$ commutators~\cite{Ernst2005, Wang2022}.
Since all Fourier components $H_{K}^{(i)}$ are comprised of only first-order ($\hat{a}$, $\hat{a}^{\dag}$) or second-order ($\hat{a}^2$, $\hat{a}^{\dag 2}$, $\hat{a}^{\dag}\hat{a}$) operators, the $K$th-order Hamiltonian is nonvanishing only when the number of first-order operators is exactly one, which occurs when the dipole tone contributes only one term~\cite{SI}.
These determine the number of tones needed at each frequency, such that the displacement $\alpha$ is
\begin{align}\label{eq:empirical_displacement_main}
    \alpha =& \beta{\left(\omega_{s}, K, \omega_{r/z}\right)} \ \Omega_{d} \ \Omega_{q}^{(K-2)/2} \  \Omega_{s}^{K/2} e^{-i\phi_s K/2} \ \\ \notag &\times\textrm{sinc}{\left( K\delta t/4\right)},
\end{align}
where $\beta$ is a function of the signal frequency $\omega_{s}$, the QHO frequency $\omega_{r/z}$, and subharmonic order $K$, independent of $\delta$~\cite{SI}.
The multimode Floquet process can be illustrated as a multiphoton transition, shown in Fig.~\ref{fig:level_diagram}(a).
The dash-dotted line represents the subharmonic excitation for a quadrupole-only configuration (ii), while the dashed lines indicate excitation for a quadrupole/dipole configuration (iii).

To verify this result and demonstrate the technique, we realize a QHO using a single trapped \ca\ ion with a radial (axial) secular frequency $\omega_r \approx 2\pi\cdot1 \textrm{ MHz}~(\omega_z= 2\pi\cdot0.7 \textrm{ MHz}$).
Doppler cooling of the ion followed by resolved-sideband cooling on the optical qubit transition prepares both the radial (axial) motional mode near the ground state with $\langle n \rangle \approx 0.01(1)$.
The dephasing time $\tau_c$ of the radial (axial) mode without applying external fields is $\approx2 \textrm{ ms}$ ($\approx 20 \textrm{ ms}$).

Subharmonic excitation can be driven on any QHO mode by applying to the trap electrodes both the signal tone at frequency $\omega_s$ in a quadrupole configuration and the probe tone at a frequency $\omega_p = \omega_s + 2\omega_{r/z}/K + \delta$ with both dipole and quadrupole components~\cite{SI}.
In what follows, we assume that all AC stark shifts of the QHO frequency from the quadrupole tones are absorbed into the resonance condition. 
Pulses are applied with a tapered-cosine envelope to mitigate off-resonant excitation~\cite{Zarantonello2019}.

\begin{figure*}
    \centering
    \includegraphics[width = 1\textwidth]{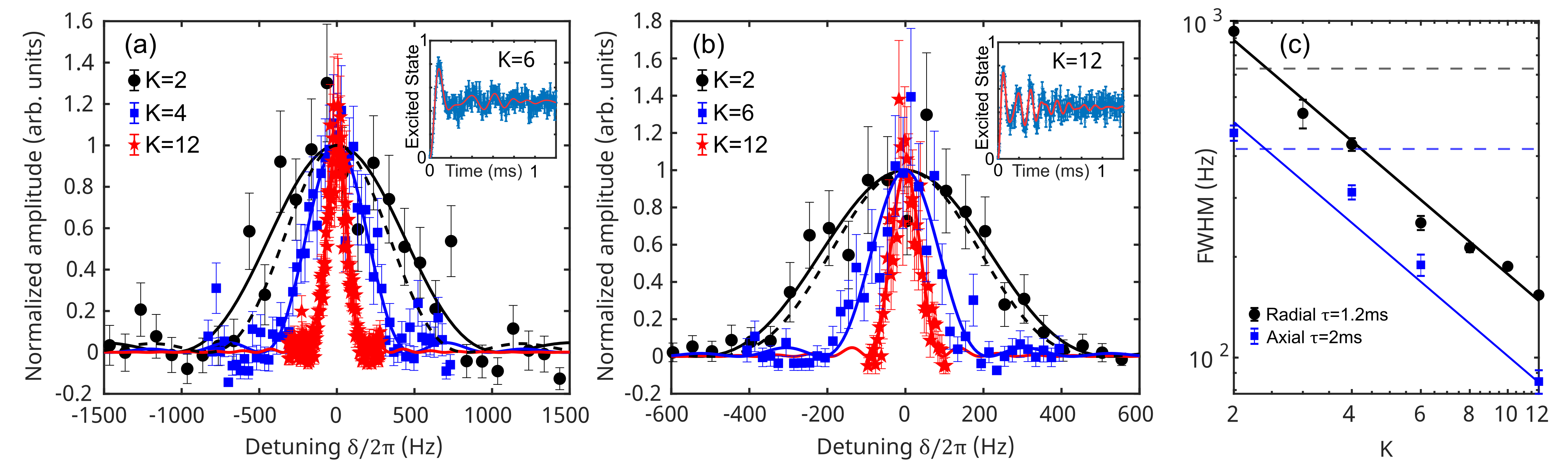}
    \caption{
    Linewidth narrowing beyond the linear Fourier transform limit (FTL) using subharmonic excitation.
    (a, b) Detection of an ``unknown" signal with subharmonic excitation at $\omega_{s}/2\pi = 70 \textrm{ MHz}$ on the radial mode (a) and $\omega_{s}/2\pi = 80 \textrm{ MHz}$ on the axial mode (b).
    The ``unknown" frequency is determined by scanning the probe detuning to maximizing the signal amplitude.
    For visibility, fit amplitudes are normalized to unity, with corresponding data rescaled by the
same factor.
    Solid lines are fits of the data to a sinc-squared lineshape. 
    Insets in (a, b): Blue sideband Rabi oscillations characterize the motional state after subharmonic excitation.
    Solid red lines are fits assuming a motional coherent state.
    (c) Dependence of linewidth on subharmonic order $K$.
    The linewidth is defined as the Full-Width at Half Maximum (FWHM).
    The solid line is a linear fit to the data with no offset.
    (a, black data in c) An effective interrogation time is $\tau = 1.2 \textrm{ ms}$ used at different values of $K$.
    (b, blue data in c) $\tau = 2.0 \textrm{ ms}$ at different values of $K$.
    $\tau$ is the pulse width of a square pulse of equivalent amplitude and area. 
    (a, b, c) Dashed lines show the FTL of the applied tapered-cosine pulse envelope.
    Error bars represent one standard error.
    }
    \label{fig:linewidth_K}
\end{figure*}

Firstly, we confirm that the dynamics of subharmonic excitation agree with our model in Eqn.~\ref{eq:empirical_displacement_main}.
We employ a motional Ramsey sequence~\cite{Wolf2019}, depicted in Fig.~\ref{fig:level_diagram}(b), to demonstrate that subharmonic excitation is a $K$-th order process.
Following cooling near the motional ground state $\ket{0}$, a resonant subharmonic excitation pulse is applied for interrogation time $\tau$ to generate a displacement $\hat{D}{\left(\alpha\right)}$.
Following free evolution for $T_{\textrm{delay}}$, a second resonant subharmonic excitation is applied with relative phase $\phi_{s}$ to generate a secondary displacement $\hat{D}{\left(\alpha'\right)}$.
For a $K$th-order process, this leads to a phase shift $K\phi_s/2$ of the displacement, confirmed in Fig.\ref{fig:level_diagram}(d) with varying order.
Further, Rabi oscillations on the blue motional sideband confirm that subharmonic excitation induces a displacement (insets in Fig.~\ref{fig:linewidth_K}~(a, b)).

We examine the line-narrowing of our protocol, relative to the linear case, by conducting spectroscopy at different subharmonic order $K$.
Specifically, we measure the frequency of a signal tone by sweeping the probe detuning $\delta$ across the subharmonic resonance (Fig.~\ref{fig:linewidth_K}~(a,b)), then measure the resultant mean phonon number $\langle n \rangle$ using the sideband ratio technique~\cite{Wu2025}.
Linewidths were extracted by fitting a sinc-squared profile $\left| \beta{\left(\omega_p\right)}\right|^2$ to the data (Fig.~\ref{fig:linewidth_K}~(c)).
At $K \geq 4$, the linewidth of the subharmonic surpasses the linear case (dashed lines in Fig.~\ref{fig:linewidth_K}~(c)), with improvement by approximately 6 times over the linear case at nonlinear $K = 12$.
To demonstrate arbitrary frequency operation, we excite the $K = 6$ subharmonic at $\omega_s/2\pi = 70$, $80$ and $200 \textrm{ MHz}$~\cite{SI}.

From Fig.~\ref{fig:linewidth_K}~(c), the linewidths narrow as $2/K$.
The $2/K$ improvement is the maximum achievable and requires a square pulse envelope; non-square pulse envelopes, e.g. Blackman or Gaussian envelopes, result in reduced narrowing due to the nonlinear dependence on amplitude.
We use a tapered-cosine pulse (Fig.~\ref{fig:level_diagram}(b)) to retain the narrowing effect while suppressing off-resonant excitation.

\begin{figure}
    \centering
    \includegraphics[width = 0.5\textwidth]{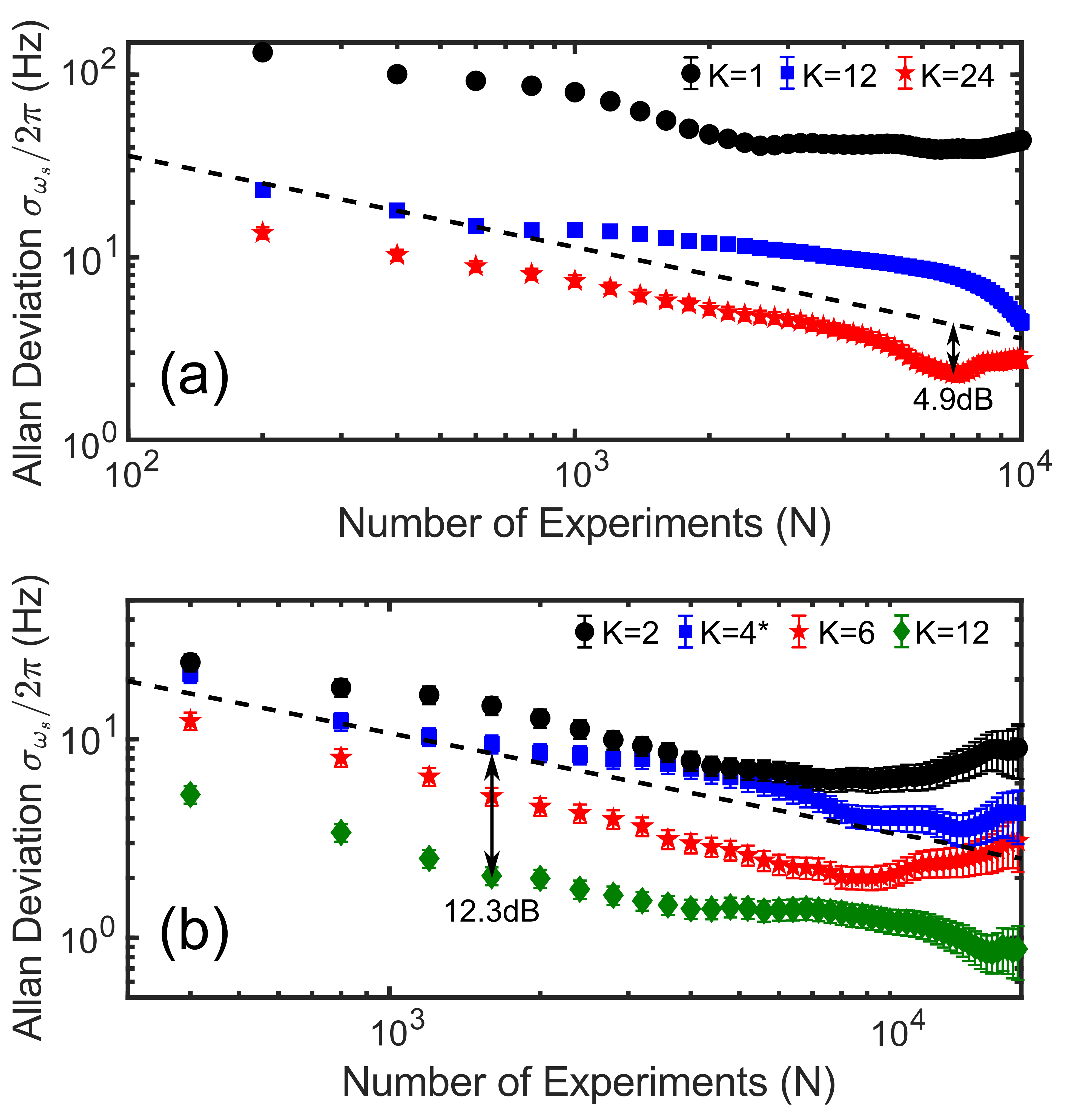}
    \caption{
    Frequency resolution of the subharmonic protocol.
    (a-b) Overlapping frequency Allan deviations for an $80~\textrm{ MHz}$ tone at different subharmonic orders $K$ on the radial (a) and axial (b) modes.
    Each experiment takes $\sim 17 \textrm{ ms}$. 
    Solid lines represent the corresponding SQL for the given interrogation time, which correspond with the SQL of the $K = 2$ subharmonic.
    (a) Subharmonic excitation is probed on the radial mode with interrogation time $\tau = 2 \textrm{ ms}$. A minimum frequency resolution $\sigma_{\omega_s}^{min}/2\pi = 2.3(2) \textrm{ Hz}$ is achieved with $K = 24$ order at $N \approx 7200$.
    A maximimum metrological gain of $4.9(8) \textrm{ dB}$ is similarly achieved with $K = 24$ at $N \approx 7000$.
    (b) Subharmonic excitation is probed on the axial mode with $\tau = 2 \textrm{ ms}$.
    A minimum frequency resolution $\sigma_{\omega_s}^{min}/2\pi = 0.85(18) \textrm{ Hz}$ is achieved with $K = 12$ at $N \approx 16000$.
    A maximum metrological gain of $12.3(9) \textrm{ dB}$ is achieved with the same $K = 12$ at $N \approx 1600$.
    Error bars represent one standard error.
    }
    \label{fig:ADEV}
\end{figure}

We next examine the \textit{useful} metrological gain relative to the best sensitivity achievable with the corresponding linear generator
using classical states, or equivalently, the SQL of K=2.
The subharmonic lineshape is first measured, as in Fig.~\ref{fig:linewidth_K}(a,b). 
Subsequent frequency measurements are made at the two half-maxima of the lineshape, then fitted to determine the center frequency.
These are used to compute the overlapping Allan ($\sigma_{\omega_s}$), shown in Figs.~\ref{fig:ADEV}~(a,b) for an $80\textrm{ MHz}$ signal at various $K$.
For comparison, the solid line represents the corresponding SQL of K=2~$(\Delta \delta^{K=2}_{SQL})$ computed from the pulse area.
A minimum frequency resolution of $0.85(18) \textrm{ Hz}$ was achieved at $N \approx 16000$ with $K = 12$ on the axial mode.
The maximum metrological gain~\cite{Pezze2018} of $20\textrm{log}_{10}(\Delta\delta^{K=2}_{SQL}/\sigma_{\omega_s}^{min})$=$12.3(9) \textrm{ dB}$ over the SQL of K=2 was realized at $N \approx 1600$ with interrogation time $\tau = 2 \textrm{ ms}$~\cite{SI}.
Likewise, a gain of $4.9(8) \textrm{ dB}$ was obtained on the radial mode with $K=24$ and $\tau = 2 \textrm{ ms}$, with a minimum resolution $2.3(2)\textrm{ Hz}$ at $N \approx 7200$.
Reduction in measurement performance on the radial mode results from increased off-resonant excitation and greater secular frequency instability.
Higher power requirements at higher orders result in a practical limit of $K=24$ ($K=12$) on the radial (axial) mode; likewise, an increase in sensing frequency results in increased power requirements. This limit is not fundamental.
These effects can be mitigated by optimizing the RF coupling efficiency to reduce the required input powers.

The achieved $12.3(9) \textrm{ dB}$ gain is comparable to that with a linear measurement on a nonclassical $\ket{n\approx 20}$ Fock state~\cite{deng2024}.
However, there is a crucial difference.
Non-classical states enhance sensitivity in parameter estimation by increasing separation between superposed states of a system, which results in greater phase accumulation for improved Fisher information~\cite{Deffner2017}.
However, non-classical states decohere faster - for example, the coherence time of a Fock state $\ket{n}$ worsens as $1/n^p$, where $p\geq 1$~\cite{Turchette2000, Myatt2000}.
This reduces \textit{usable} measurement times to largely eliminate any metrological gain over the SQL in the context of ultimate precision.
Unlike commonly used active approaches that can mitigate the reduced coherence time associated with nonclassical states—such as decoherence-free subspaces developed for spin systems \cite{Bruzewicz2019} or bosonic quantum error-correction codes such as binomial codes \cite{Michael2016}, our protocol requires only classical input states and derives metrological gain from a \textit{nonlinear} interaction. As a result, it passively avoids the excess decoherence typically introduced by nonclassical probe states, thereby improving the \textit{ultimate} precision of frequency sensing for bosonic systems such as motional modes.

Finally, a minimum frequency uncertainty $\sigma_{\omega_s}^{min}/2\pi = 0.56(32) \textrm{ Hz}$ was achieved at $80 \textrm{ MHz}$ with $\tau = 5 \textrm{ ms}$ on the axial mode~\cite{SI}.
To our knowledge, this surpasses the precision of previous frequency measurements in the radio-frequency band using QHO-based system and sets a new state-of-the-art for these measurements.
This measurement could easily be improved by approximately 4 times by improving damage thresholds of system electronics, with further gains achievable by extending motional coherence times.

In summary, we have demonstrated high-resolution frequency measurements using a nonlinear generator that surpasses the SQL of corresponding linear generator, enabled by driving Floquet-engineered motional Raman subharmonics.
This technique is adaptable to platforms supporting Raman transitions, including nitrogen-vacancy centers, superconducting circuits, and neutral atom systems.
With a stable reference, this method refines electromagnetic-field frequency sensing across the RF, microwave, and even optical domains.
Importantly, our technique requires only classical input states, eliminating the excess decoherence that limits measurements with nonclassical states to improve \textit{ultimate} precisions for frequency sensing.
Despite this, our technique remains compatible with nonclassical states of motion~\cite{Wu2025} (e.g. Fock states~\cite{Wolf2019, Wolf2021, McCormick2019, deng2024}) for short-timescale metrological gain to surpass the SQL of the nonlinear generator.

\section{Acknowledgements}
We thank Dylan Kawashiri for building a computer cluster used to simulate these results.
This work was supported by NSF (PHY-2110421 and OMA-2016245), AFOSR (130427-5114546), and ARO (W911NF-19-1-0297).

\section{Competing interests}
The authors declare no competing interests.

\section{Data availability}
The data supporting the findings of this article are openly available~\cite{Note1}.

%


\appendix
\onecolumngrid


\section{Supplementary Information}
\subsection{Appendix A: Theoretical Analysis}
\begin{figure}[h!]
    \centering
    \includegraphics[width = 0.8\textwidth]{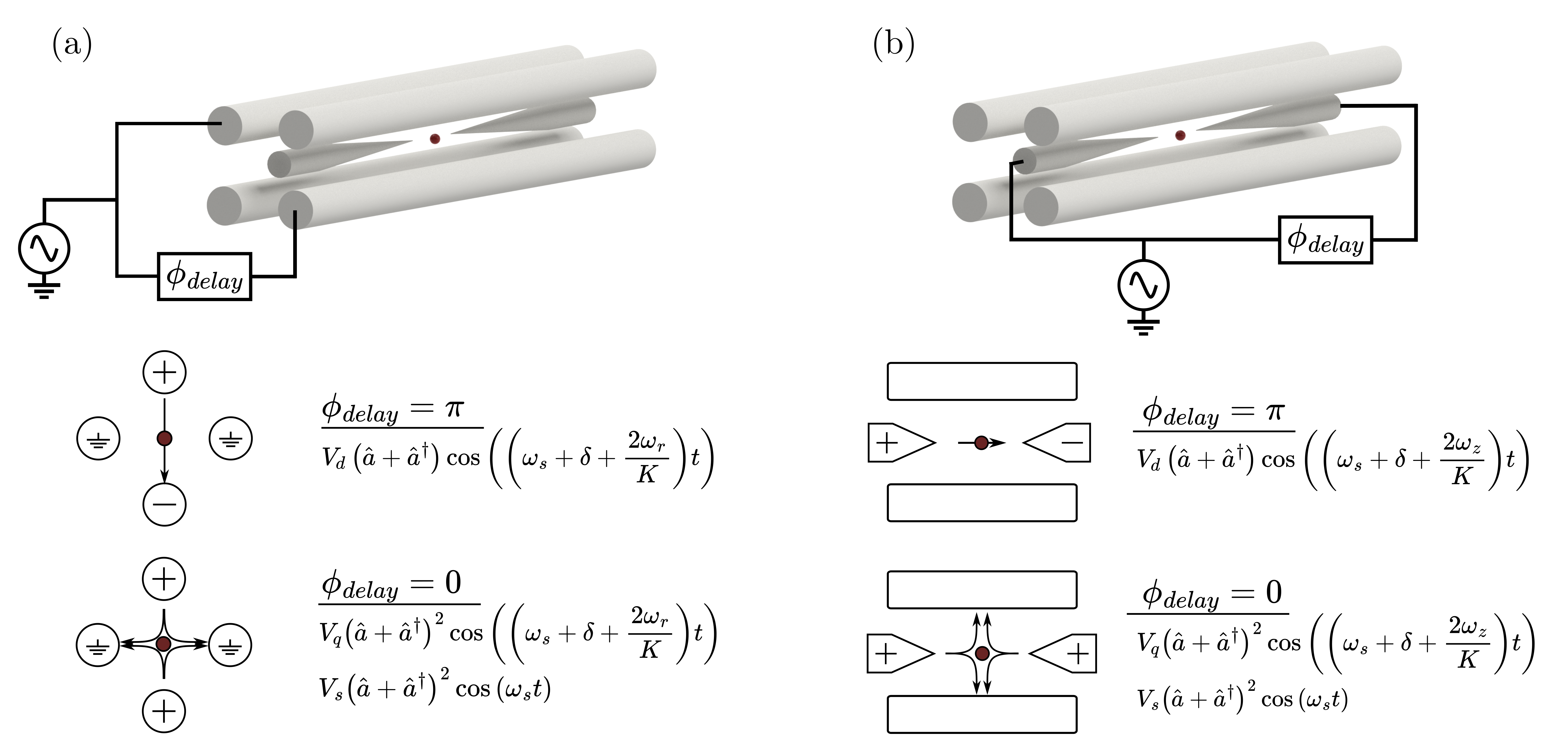}
    \caption{
    Implementation of subharmonic excitation on the radial and axial modes, respectively.
    (a) Subharmonic excitation is applied to excite the radial motional mode by coupling an oscillating voltage onto a pair of opposing trap electrodes.
    (b) The axial motional mode can be excited by instead coupling the voltage onto the axial endcaps of the trap.
    A dipole field is generated for a relative phase $\phi_{\textrm{delay}} = \pi$ between the voltages coupled to the electrodes, while a quadrupole field is generated if the coupled voltages are exactly in phase (i.e. $\phi_{\textrm{delay}} = 0$).
       }
    \label{fig:scheme}
\end{figure}
In this section, we derive the dynamics of motional excitation on a subharmonic resonance within the framework of multimode Floquet theory~\cite{Ernst2005, Wang2022}.
Consider an unknown signal tone $\omega_s$ to be detected, which is coupled to the trapping electrodes in a quadrupole configuration.
Additionally, a probe tone $\omega_p$ is applied in a combination of dipole and quadrupole configurations as shown in Fig.~\ref{fig:scheme}.
The resulting Hamiltonian representing the motional state of the harmonic oscillator and electric field coupling can be is: 
\begin{align}
    \frac{\hat{H}}{\hbar} = \omega_{r/z} \hat{a}^{\dagger} \hat{a}
    &+ \Omega_{q} \left( \hat{a} + \hat{a}^{\dagger} \right)^2
    \cos{ \left( \omega_p t \right)} \notag \\
    &+ \Omega_{d} \left( \hat{a}+\hat{a}^\dagger \right) \cos{ \left( \omega_p t \right)}\notag \\
    &+ \Omega_{s} \left( \hat{a} + \hat{a}^{\dagger} \right)^2
    \cos{ \left( \omega_{s} t + \phi_{s} \right)}, 
    \label{eq:hamiltonian_base_full}
\end{align}
where $\omega_p=\omega_s+\delta+2\omega_{r/z}/K$, $\Omega_{q} = eV_{q} \eta^2/\hbar$, $\Omega_{d} =  eV_{d} \eta/\hbar$, $\Omega_{s} = eV_{s}\eta^2/\hbar$, $\eta=x_0/r_0$ is effective Lamb-Dicke parameter, $r_o$ is the ion-electrode distance,
$x_o = \sqrt{\hbar/(2m\omega_{r/z})}$ is the zero-point wavefunction of the QHO,
$e$ is the electron charge, and $\delta$ is the detuning.
Without loss of generality, we ignore the phase of the probe tones and use $\omega$ to represent $\omega_{r/z}$.

Unlike previous schemes~\cite{SUDAKOV2000, Zhao2002, COLLINGS2000, Tommaseo2003} which have exclusively employed quadrupole tones, the present approach applies an additional probe tone in the dipole configuration for several reasons:
\begin{itemize}
    \item  The effective Lamb-Dicke parameter $\eta$ associated with a dipole tone is of a lower order than that of quadrupole tones, which reduces the power threshold required to excite higher-order subharmonics.
    \item Application of a dipole tone induces motional excitation that generates a coherent state rather than a squeezed state.
    Coherent states operate within a linear regime compared to squeezed states, making them more straightforward to manipulate compared to a squeezed state.
    \item Measurement of a coherent state provides higher Fisher information, leading to enhanced sensitivity in parameter estimation, such as for frequency measurements.
\end{itemize}

Transforming into the interaction picture with respect to $\omega \hat{a}^{\dagger} \hat{a}$, the Hamiltonian \eqref{eq:hamiltonian_base_full} becomes:
\begin{align}
    \frac{\hat{H_I}}{\hbar} = &+\Omega_{q}/2 \left( \hat{a}^2 e^{-i(2\omega+\omega_p)t} + \hat{a}^{\dagger 2} e^{
    i(2\omega+\omega_p)t} \right) \notag\\
    &+\Omega_{q}/2 \left( \hat{a}^2 e^{i(\omega_p-2\omega)t} + \hat{a}^{\dagger 2} e^{-i(\omega_p-2\omega)t} \right) \notag\\
    &+\Omega_{q}/2 \left(2\hat{a}^{\dagger} \hat{a}+1 \right) \left(e^{i \omega_p t}+e^{-i \omega_p t} \right) \notag\\
    &+\Omega_{d}/2 \left(\hat{a} e^{-i(\omega+\omega_p)t}+\hat{a}^{\dagger} e^{i(\omega+\omega_p)t} \right) \notag\\
    &+\Omega_{d}/2 \left(\hat{a} e^{i(\omega_p-\omega)t}+\hat{a}^{\dagger} e^{-i(\omega_p-\omega)t} \right) \notag\\
    &+\Omega_{s}/2 \left( \hat{a}^2 e^{-i(2\omega+\omega_s+\phi_s)t}+(\hat{a}^{\dagger})^2e^{
    i(2\omega+\omega_s+\phi_s)t} \right) \notag\\
    &+\Omega_{s}/2 \left( \hat{a}^2 e^{i(\omega_s-2\omega+\phi_s)t} + \hat{a}^{\dagger 2} e^{
    -i(\omega_s-2\omega+\phi_s)t} \right) \notag\\
    &+\Omega_{s}/2 \left(2\hat{a}^{\dagger} \hat{a}+1 \right) \left(e^{i (\omega_s t+\phi_s)}+e^{-i( \omega_s t+\phi_s)} \right)
    \label{eq:hamiltonian_interaction_full}
\end{align}
Within the framework of multimode Floquet theory~\cite{Ernst2005,Wang2022}, the effective Hamiltonian $\hat{H_I}$ can be represented as an eight-mode Floquet Hamiltonian:
\begin{align}
    \bar{H_I}{\left(t\right)} = & \sum_{\ev{c_1,c_2,c_3,c_4,c_5,c_6,c_7,c_8}} \left( H_{(1)}^{(c_1,c_2,c_3,c_4,c_5,c_6,c_7,c_8)} + H_{(2)}^{(c_1,c_2,c_3,c_4,c_5,c_6,c_7,c_8)} + \cdots + H_{(K)}^{(c_1,c_2,c_3,c_4,c_5,c_6,c_7,c_8)} + \cdots \right) \notag\\
    & \cdot \mathrm{e}^{-i (\sum_{k=1}^8 c_k\omega_k) t-i(c_6+c_7+c_8)\phi_s},
    \label{eq:floquet_hamiltonian2}
\end{align}
where for $H_{(K)}^{(c_1,c_2,c_3,c_4,c_5,c_6,c_7,c_8)}$, the subscript $K$ denotes the corresponding order of subharmonic excitation and the number of tones involved in the nonlinear process,
while the superscript values $(c_1,c_2,c_3,c_4,c_5,c_6,c_7,c_8)$ represent the eight selectable tones -- 
$\omega_1=\omega_p+2\omega$,
$\omega_2=\omega_p-2\omega$,
$\omega_3=\omega_p$,
$\omega_4=\omega_p+\omega$,
$\omega_5=\omega_p-\omega$,
$\omega_6=\omega_s+2\omega$,
$\omega_7=\omega_s-2\omega$,
and $\omega_8=\omega_s$.

The Fourier components corresponding to the eight selectable tones are given as follows:
\begin{align}
   & H^{(0,-1,0,0,0,0,0,0)}=H^{(1,0,0,0,0,0,0,0)}=\Omega_{q}/2 (\hat{a}^{\dagger})^2,\notag\\
   & H^{(0,1,0,0,0,0,0,0)}=H^{(-1,0,0,0,0,0,0,0)}=\Omega_{q}/2 (\hat{a})^2,\notag\\
   & H^{(0,0,-1,0,0,0,0)}=H^{(0,0,1,0,0,0,0)}=\Omega_{q}/2 \left(2\hat{a}^{\dagger} \hat{a}+1 \right),\notag\\
   & H^{(0,0,0,0,-1,0,0,0)}=H^{(0,0,0,1,0,0,0,0)}=\Omega_{d}/2 (\hat{a}^{\dagger}),\notag\\
   & H^{(0,0,0,0,1,0,0,0)}=H^{(0,0,0,-1,0,0,0,0)}=\Omega_{d}/2 (\hat{a}),\notag\\
   & H^{(0,0,0,0,0,0,-1,0)}=H^{(0,0,0,0,0,1,0,0)}=\Omega_{s}/2 (\hat{a}^{\dagger})^2,\notag\\
   & H^{(0,0,0,0,0,0,1,0)}=H^{(0,0,0,0,0,-1,0,0)}=\Omega_{s}/2 (\hat{a})^2,\notag\\
   & H^{(0,0,0,0,0,0,0,-1)}=H^{(0,0,0,0,0,0,0,1)}=\Omega_{s}/2 \left(2\hat{a}^{\dagger} \hat{a}+1 \right),
    \label{eq:floquet_components}
\end{align}

When the resonance condition $\sum_{k=1}^8 c_k\omega_k=0$ is satisfied, a higher-order subharmonic is excited, generating a motional displacement.
The resonance condition can be reduced to $(\sum_{k=1}^8 c_k)\omega_s+((\sum_{k=1}^5 c_k)2/K+(2c_1-2c_2+c_4-c_5+2c_6-2c_7))\omega+(\sum_{k=1}^5 c_k)\delta=0$ to yield the constraints:
\begin{align}
    c_1+c_2+c_3+c_4+c_5+c_6+c_7+c_8=0, \notag\\
    (2c_1-2c_2+c_4-c_5+2c_6-2c_7)=-1, \notag\\
    c_1+c_2+c_3+c_4+c_5=K/2.
    \label{eq:floquet_res_cond}
\end{align}
These constraints can be intuitively understood as a multi-phonon Raman transition. To generate a displacement resonant with the secular frequency $\omega$, the number of probe tones must match the number of signal tones to cancel the carrier frequency $\omega_s$.

Creating a general analytical expression for an arbitrary $K$th-order Hamiltonian that includes all parameters is nontrivial.
Instead, we here derive an empirical formula.

The $K$th-order Hamiltonian $H_{(K)}^{(c_1,c_2,c_3,c_4,c_5,c_6,c_7,c_8)}$ involves $K$-1 commutation operations~\cite{Ernst2005,Wang2022}.
From \eqref{eq:floquet_components}, all Fourier components $H_{(K)}^{(c_1,c_2,c_3,c_4,c_5,c_6,c_7,c_8)}$ are comprised of only first-order ($\hat{a}$, $\hat{a}^{\dag}$) or second-order ($\hat{a}^2$, $\hat{a}^{\dag 2}$, $\hat{a}^{\dag}\hat{a}$) ladder operators.
The commutation between ladder operators is nonvanishing only when the number of first-order ladder operators is exactly one, which occurs only if the dipole tone contributes only one of $K$ terms in the Hamiltonian.
Together with the resonance constraints, we arrive at an empirical formula for subharmonic driving of a coherent state:
\begin{align}
    \alpha = \beta(\omega_{s}, K, \omega_{r/z}) \ \Omega_{ d} \ \Omega_{q}^{(K-2)/2} \  \Omega_{s}^{K/2}e^{-i\phi_s K/2},
    \label{eq:empirical_displacement_SI}
\end{align}
where $\beta$ is a function of the signal frequency $\omega_{s}$, the secular frequency $\omega$, and the subharmonic order $K$, and can be numerically calculated from QuTip simulations~\cite{JOHANSSON2012,JOHANSSON2013}.

To validate the Floquet theory for the subharmonic excitation, we consider the case of $K$=2 as an example.
The second-order term for eight-mode Floquet theory can be expressed as:
\begin{align}
    H_{(2)}^{(c_1,c_2,c_3,c_4,c_5,c_6,c_7,c_8)}=-\frac{1}{2}\sum_{\substack{c_1',c_2',c_3',c_4',\\c_5',c_6',c_7',c_8'}}\frac{[H^{(c_1,c_2,c_3,c_4,c_5,c_6,c_7,c_8)},H^{(c_1'-c_1,c_2'-c_2,c_3'-c_3,c_4'-c_4,c_5'-c_5,c_6'-c_6,c_7'-c_7,c_8'-c_8)}]}{\sum_{k=1}^8 c_k\omega_k},
    \label{eq:floquet_ex_2nd_hamil}
\end{align}
Evaluation of the sum \eqref{eq:floquet_ex_2nd_hamil} gives:
\begin{align}
    H_{(2)}= & \ 2 \left( H_{(2)}^{(0,0,0,1,0,-1,0,0)} + H_{(2)}^{(0,0,0,-1,0,1,0,0)} + H_{(2)}^{(0,0,0,0,1,0,0,-1)} + H_{(2)}^{(0,0,0,0,-1,0,0,1)} \right) \notag\\
    = &- \bigg( \frac{\left[H^{(0,0,0,0,0,-1,0,0)}, H^{(0,0,0,1,0,0,0,0)}\right]}{\omega_p + \omega} + \frac{\left[H^{(0,0,0,0,0,1,0,0)}, H^{(0,0,0,-1,0,0,0,0)}\right]}{-(\omega_p + \omega)} \notag\\
    & +\frac{\left[H^{(0,0,0,0,0,0,0,-1)},H^{(0,0,0,0,1,0,0,0)}\right]}{\omega_p-\omega}
    +\frac{\left[H^{(0,0,0,0,0,0,0,1)},H^{(0,0,0,0,-1,0,0,0)}\right]}{-(\omega_p-\omega)} \bigg) \notag\\
    = & -\frac{\Omega_s\Omega_{d}}{4(\omega_p+\omega)} \left[\hat{a}^{2},\hat{a}^{\dag}\right]
    +\frac{\Omega_s\Omega_{d}}{4(\omega_p+\omega)} \left[\hat{a}^{\dag 2},\hat{a}\right]
    -\frac{\Omega_s\Omega_{d}}{4(\omega_p-\omega)} \left[2\hat{a}^{\dag}\hat{a},\hat{a}\right]
    +\frac{\Omega_s\Omega_{d}}{4(\omega_p-\omega)} \left[2\hat{a}^{\dag} \hat{a},\hat{a}^{\dag} \right] \notag\\
    = & \:  \frac{\omega \Omega_s\Omega_{d}}{\omega_p^2 - \omega^2}(\hat{a}^{\dag} + \hat{a}),
    \label{eq:floquet_ex_final_hamil}
\end{align}
which is consistent with the result obtained from the Magnus expansion~\cite{Wu2025}.

The QHO squeezing operator $\hat{S}(r)$, generated via parametric excitation, can also be analyzed using Floquet theory.
Similarly, application of a single quadrupole tone at frequency $\omega_{p}$ results in a Hamiltonian:
\begin{align}
    \frac{\hat{H}}{\hbar} = \omega \hat{a}^{\dagger} \hat{a}
    &+ \Omega_{q} \left( \hat{a} + \hat{a}^{\dagger} \right)^2
    \cos{ \left( \omega_p t \right)}, 
    \label{eq:hamiltonian_parametric}
\end{align}
where $\omega_p=2\omega/K+\delta$ and $\Omega_{q} = eV_{q} \eta^2/\hbar$.

Transforming into the interaction picture with respect to $\omega \hat{a}^{\dagger} \hat{a}$, the Hamiltonian \eqref{eq:hamiltonian_parametric} becomes:
\begin{align}
    \frac{\hat{H}_{I}}{\hbar} = &\frac{\Omega_{q}}{2} \left( \hat{a}^2 e^{-i\left(2\omega+\omega_p\right)t} + \hat{a}^{\dagger 2} e^{i\left(2\omega+\omega_p\right)t} \right)\notag\\
    + &\frac{\Omega_{q}}{2} \left( \hat{a}^2 e^{i\left(\omega_p-2\omega\right)t} + \hat{a}^{\dagger 2} e^{
    -i\left(\omega_p-2\omega\right)t} \right)\notag\\
    + &\frac{\Omega_{q}}{2} \left(2\hat{a}^{\dagger} \hat{a}+1 \right) \left(e^{i \omega_p t}+e^{-i \omega_p t} \right)\label{eq:hamiltonian_interaction_parametric_full}
\end{align}
The Hamiltonian can be treated as a sum of three Hermitian components, each composed of a factor $\Omega_i$, ladder operators, and a time dependence $e^{\pm \imath \omega_i t}$, where
$\{\Omega_1 = \Omega_q, \omega_1=(2/K+2)\omega\}$,
$\{\Omega_2 = \Omega_q, \omega_2=(2/K-2)\omega\}$ and
$\{\Omega_3 = \Omega_q, \omega_3=2\omega/K\}$.

Using Eqs.~(A16) and (A17) in Ref.~\cite{Wang2022}, the $K^{\textrm{th}}$-order term of the Hamiltonian can be approximately expressed as $\hat{H} =  \Omega^K/\left( \omega/2 \right)^{K-1} \left( \hat{a}^2 + \hat{a}^{\dagger 2} \right)$, creating a squeezing operator $\hat{S}{\left( r \right)}$ where $r \propto \Omega^K/(\omega / 2)^{K-1}$.

\subsection{Appendix A': Time-dependence of the $K$-th order Interaction}

From Eqn.~\ref{eq:floquet_components}, significant excitation occurs when near a motional Raman subharmonic resonance, i.e.
\begin{align}
    \sum_{k=1}^{8}{c_k\omega_k} = \gamma,
    \label{eq:tmp_alt_res_1}
\end{align}
where $\gamma$ is some small detuning from resonance and $\gamma \approx \delta \ll \omega \ll \omega_{s}, \omega_{p}$.
This resonance condition can be expanded
\begin{align}
    \left(\sum_{k=1}^{8}{c_{k}} \right) \omega_{s} +
    \left( \left( \sum_{k=1}^5 c_k \right) \frac{2}{K} + \left(2c_1-2c_2+c_4-c_5+2c_6-2c_7 \right) \right) \omega +
    \left( \sum_{k=1}^5 c_{k} \right) \delta = \gamma,
    \label{eq:tmp_alt_res_2}
\end{align}
which in turn yields the constraints
\begin{align}
    c_1+c_2+c_3+c_4+c_5+c_6+c_7+c_8=0, \notag\\
    (2c_1-2c_2+c_4-c_5+2c_6-2c_7)=-1, \notag\\
    c_1+c_2+c_3+c_4+c_5=K/2.
    \label{eq:tmp_alt_res_3}
\end{align}

Collecting terms of order $O{\left( \delta\right)}$, the resonance condition \eqref{eq:tmp_alt_res_1} is only fulfilled when $\left( \sum_{k=1}^5 c_{k} \right) \delta = \gamma$ since $\gamma \approx \delta \ll \omega \ll \omega_{s}, \omega_{p}$.
Use of Eqn.~\ref{eq:tmp_alt_res_2} with Eqn.~\ref{eq:tmp_alt_res_3} thus gives the result
\begin{align}
    \frac{K \delta}{2} = \gamma.
    \label{eq:tmp_alt_res_4}
\end{align}
The time-dependence of the $K$-th order Floquet Hamiltonian \eqref{eq:floquet_hamiltonian2} is thus 
\begin{align}
    \mathrm{e}^{i\gamma t} = \mathrm{e}^{-i K\delta t/2  -i \left(c_6+c_7+c_8\right) \phi_s},
    \label{eq:tmp_alt_res_5}
\end{align}

More generally, for near-resonant detunings, the $K$-th order Floquet Hamiltonian Eqn.~\ref{eq:floquet_hamiltonian2} is similarly
\begin{align}
    \bar{H}_{I}{\left(t\right)} = & \sum_{\ev{c_1,c_2,c_3,c_4,c_5,c_6,c_7,c_8}}
    \left( H_{(1)}^{(c_1,c_2,c_3,c_4,c_5,c_6,c_7,c_8)} + \cdots + 
    H_{(K)}^{(c_1,c_2,c_3,c_4,c_5,c_6,c_7,c_8)} + \cdots \right) \notag \\
    & \cdot \mathrm{e}^{-i K\delta t/2  -i \left(c_6+c_7+c_8\right) \phi_s}+H.C.,    \label{eq:floquet_hamiltonian3}
\end{align}

Again, since we know the Hamiltonian can be written in a form of a displacement operator, Eqn.~\ref{eq:floquet_hamiltonian3} can be simplified to:
\begin{align}
    \bar{H_I}{\left(t\right)} = & \ C \cdot \hat{a}\cdot \mathrm{e}^{-i K/2 \delta t-i(c_6+c_7+c_8)\phi_s}+ \ C^* \cdot \hat{a}^+\cdot \mathrm{e}^{i K/2 \delta t+i(c_6+c_7+c_8)\phi_s},    \label{eq:floquet_hamiltonian4}
\end{align}
where $C$ is a $\delta$-independent amplitude. Following Eqns.~(14-20) of Ref.~\cite{Lee2005}, Eqn.~\ref{eq:floquet_hamiltonian4} takes the form of an oscillating displacement where
\begin{align}
    \alpha = C \cdot\textrm{sinc}{\left( K\delta t/4\right)}.
\end{align}
Together with Eqn.~\ref{eq:empirical_displacement_SI}, the displacement $\alpha$ can be made explicitly as
\begin{align}
    \alpha = \beta{\left(\omega_{s}, K, \omega_{r/z}\right)} \ \Omega_{d} \ \Omega_{q}^{(K-2)/2} \  \Omega_{s}^{K/2} e^{-i\phi_s K/2} \ \textrm{sinc}{\left( K\delta t/4\right)},
\end{align}
which is shown in Eqn.~\ref{eq:empirical_displacement_main} of the main text.

\subsection{Appendix B: QuTiP Simulations}

\begin{figure}[h!]
    \centering
    \includegraphics[width = 0.8\textwidth]{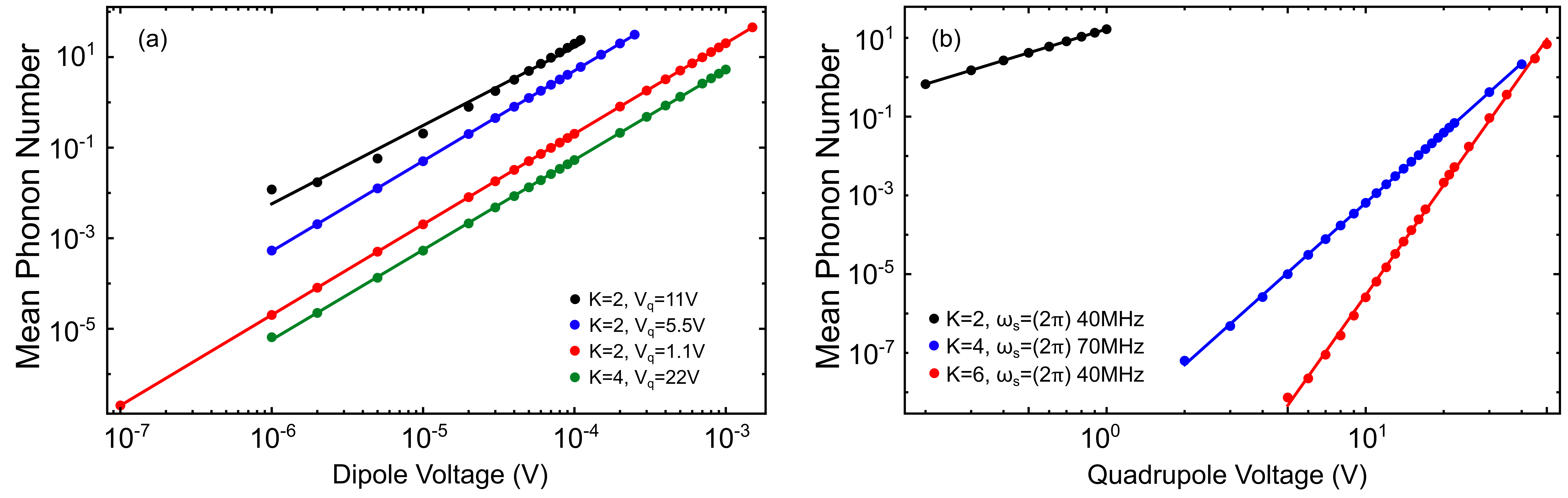}
    \caption{
        (a, b) Dependence of the subharmonic excitation on the applied dipole voltage $V_{d}$ and the applied quadrupole voltages $V_{s}, V_{q}$, respectively.
        Results are simulated using a $\omega = 1 \textrm{ MHz}$ secular frequency.
        A linear model $y = mx + b$ is fitted to the log-log transform of the results to extract the power-law dependence at each subharmonic order.
        (a) Simulated results at all subharmonic orders exhibit an approximately quadratic dependence of mean phonon number on dipole voltage $V_{d}$, equivalent to a linear dependence of displacement on $V_{d}$.
        $\omega_s/2\pi=40$~MHz.
        (b) Simulated results exhibit a power-law scaling of the mean phonon number on $V_{q}$, using a $ {V_{d}=1}\textrm{ mV}$, that agrees well with the predicted scaling of $2(K-1)$. 
    }
    \label{fig:sim_qutip_dipole_scaling}
\end{figure}

Simulations were conducted in QuTiP by time-evolving the initial state under the influence of the subharmonic excitation Hamiltonian \eqref{eq:hamiltonian_base_full} via the time-dependent Schrödinger equation.
Constituent data points in Figs.~\ref{fig:sim_qutip_dipole_scaling}(a-b)
were obtained by simulating the system at different detunings $\delta$ to extract a lineshape, which was fitted using a sinc-squared lineshape to extract the maximum mean phonon number, linewidth, and center frequency.
This procedure was repeated at different values of the target variable to extract the dependence on the target variable.
The subharmonic excitation Hamiltonian was pulse-shaped in simulation using a tapered-cosine window with $100\mu\textrm{s}$ rise and fall times and a total length of $1.2 \textrm{ms}$.
The basis size was varied for each data point to reduce simulation overheads, though confirmed to be sufficiently large to contain all simulation dynamics.
The Fock state populations and Wigner functions were extracted for each simulation to confirm that a coherent state was produced.

The mean phonon number scales as $V_{d}^{2}$ for all orders of subharmonic excitation, where $V_{d}$ is the dipole voltage, indicating an overall linear dependence of the displacement on $V_{d}$, in agreement with the theoretical analysis and the empirical formulae \eqref{eq:empirical_displacement_main}, \eqref{eq:empirical_displacement_SI}.
Simulations instead varying the quadrupole voltages $V_{s} \textrm{, } V_{q}$ show that the resultant phonon number scales as $\left( V_{s} V_{q}\right)^{2\left(K - 1\right)}$, indicating that the displacement scales as $K-1$, again in agreement with the theoretical analysis and the empirical formula \eqref{eq:empirical_displacement_SI}.

\begin{figure}[h!]
    \centering
    \includegraphics[width = 0.4\textwidth]{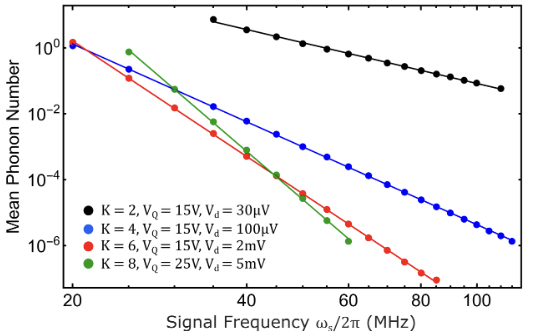}
    \caption{
        Dependence of the subharmonic excitation on the signal tone frequency $\omega_{s}$.
        Results are simulated using a $\omega = 1 \textrm{ MHz}$ secular frequency.
        A linear model $y = mx + b$ is fitted to the log-log transform of the results to extract the power-law dependence at each subharmonic order.
        Simulated results exhibit a power-law scaling of the mean phonon number on $\omega_{s}$ in good agreement with the predicted scaling of $\langle n \rangle \propto \omega_{s}^{2K}$, or equivalently, $\alpha \propto 1/\omega_{s}^{K}$.}
    \label{fig:sim_qutip_raman_scaling}
\end{figure}

Simulations in Fig.~\ref{fig:sim_qutip_raman_scaling} examine the dependence of the mean phonon number $\langle n \rangle$ on the signal frequency $\omega_{s}$.
These results demonstrate that $\langle n \rangle \propto 1/\omega_{s}^{2K}$, indicating that the displacement scales as $\alpha \propto 1/\omega_{s}^{K}$.

\subsection{Appendix C: The Standard Quantum Limit (SQL) - Frequency Sensing}

We here derive a general expression for the frequency uncertainty ($\Delta \delta$) for the $K$-th order subharmonic.
Since only the $\delta$-dependent term is relevant for deriving the uncertainty ($\Delta\delta$), Eqn.~\ref{eq:empirical_displacement_main} in the main text can be simplified as:
\begin{align}
   \alpha={\lvert \Tilde{a} \rvert}
   \frac{\sin{\left(K\delta \tau/4\right)}}{\left(K\delta \tau/4\right)}
    \label{eq:uncertainy_K},
\end{align}
where $\tilde{a}$ represents the product of all prefactors multiplying the \textit{Sinc} function in Eqn.~3 and corresponds to the maximum displacement in phase space,  $\tau$ is the interrogation time, and $N$ is the number of experiments. $\tilde{a}$ is typically fixed at approximately 0.6 across different experiments and values of $K$.

The frequency uncertainty $\Delta\delta$ can be obtained through standard error propagation:
\begin{align}
    \Delta\delta = \frac{\Delta \alpha}{\pdv{\alpha}{\delta}} =
    \frac{4\Delta\alpha}{K\tau}
    \frac{\left(K\delta \tau/4 \right)^2}{(K\delta \tau/4 \cos{\left(K\delta \tau/4\right)} -  \sin{\left(K\delta \tau/4\right))}} \frac{1}{ \lvert\Tilde{a}\rvert\sqrt{N},}
    \label{eq:sql_f_SI}
\end{align}
where $\Delta\alpha$ denotes the uncertainty in 
$\alpha$ for a single measurement. Its value depends on the Fisher information associated with the measurement technique and is ultimately bounded by 0.5 for a classical state. 
The actual frequency uncertainty is determined from an Allan deviation analysis, denoted $\sigma_{\omega_s}$. 
In this approach, two operating points close to the half-maximum of the signal are used, which optimally balance sensitivity and signal-to-noise ratio. Under these conditions, $\textrm{sinc}{\left( \frac{K}{4}\delta \tau \right)} = 1/\sqrt{2}$
Eqn.~\ref{eq:sql_f_SI} thus becomes:
\begin{align}
   \Delta\delta \approx 2\pi
   \frac{4\Delta\alpha}{K}
   \frac{0.418}{\lvert\Tilde{a}\rvert \tau \sqrt{N}},
    \label{eq:sql_f_SI_adev}
\end{align}
Increasing $K$ thus leads to a smaller uncertainty of $\delta$.

When $\Delta \alpha$ reaches the shot-noise limit, i.e. $\Delta \alpha = 1/2$, the frequency uncertainty $\Delta\delta_{SQL}$ corresponding to the SQL of Kth-order nonlinear generator is therefore:
\begin{align}
   \Delta\delta_{SQL} \approx 2\pi
   \frac{2}{K}
   \frac{0.418}{\lvert\Tilde{a}\rvert \tau \sqrt{N}}.
    \label{eq:sql_f_SI3}
\end{align}

To benchmark our method, we calculate the best sensitivity achievable with the corresponding linear generator using classical states, or equivalently, the SQL associated with that linear generator.
This SQL corresponds to the special case of $K=2$, which can be obtained directly from Eqn.~\ref{eq:sql_f_SI_adev} and the expression is shown below:
\begin{align}
    \Delta\delta^{K=2}_{SQL} \approx 2\pi \frac{0.418}{\lvert\Tilde{a}\rvert \tau \sqrt{N}}.
    \label{eq:sql_f_SI_adev2}
\end{align}

\subsection{Appendix D: Wideband Sensing with Subharmonic Excitation}

\begin{figure}[h!]
    \centering
    \includegraphics[width = 0.8\textwidth]{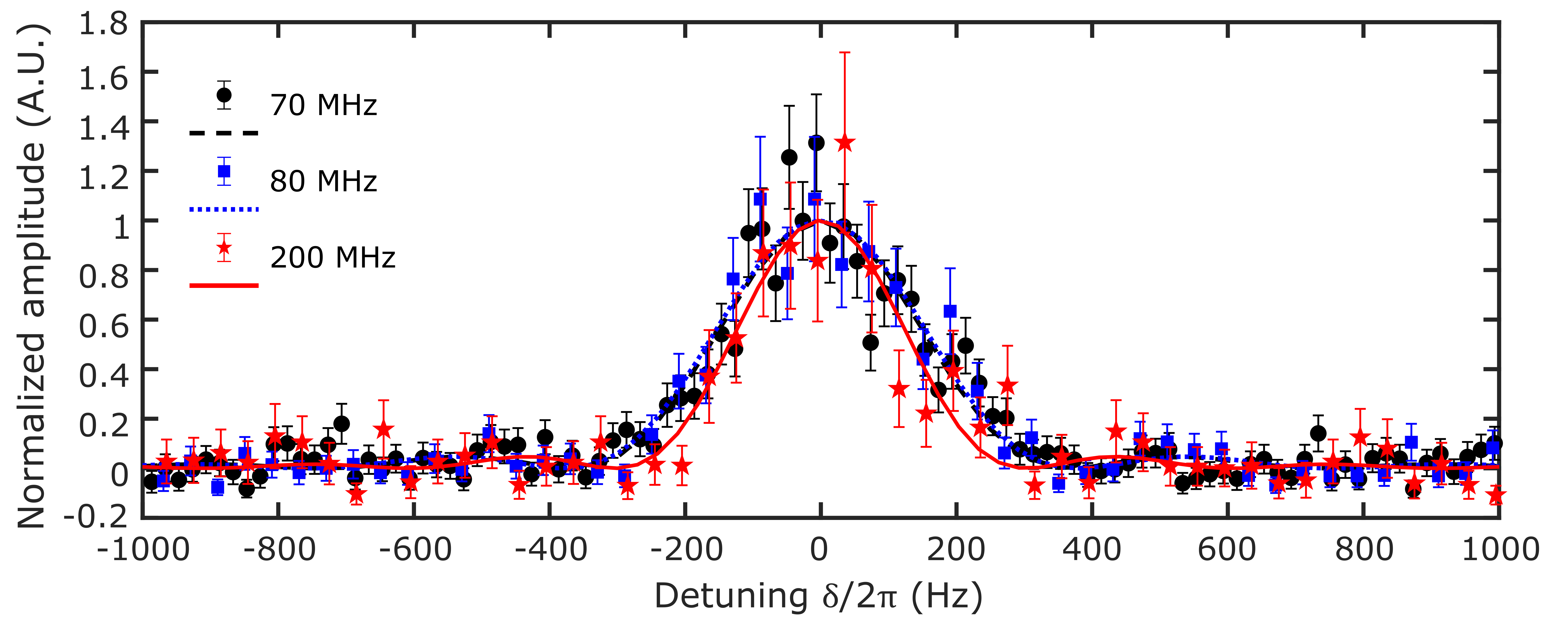}
    \caption{
    $K$=6 subharmonic excitation at different signal frequencies~($\omega_s/2\pi$)=(70, 80 and 200~MHz). 
    Lines are fits to the data using a sinc-squared profile.
    All data are normalized by the peak amplitudes of their fit.
    The interrogation time is fixed at $\tau = 1.2 \textrm{ ms}$.
    Error bars represent one standard error.
    }
    \label{fig:diff_carriers}
\end{figure}

To confirm the wide frequency operation of our protocol, we characterized the $K$=6 subharmonic resonance at three different signal frequencies $\omega_s/2\pi$= 70, 80 and 200 MHz (Fig.~\ref{fig:diff_carriers}).
Traces are normalized by maximum excitation and exhibit highly overlapping lineshapes, making our protocol uniquely suited to the challenges of wideband quantum sensing.
The sensing with the $K=6$ subharmonic at $\omega_{s} = 200 \textrm{ MHz}$ required an RF power of roughly $33 \textrm{ dBm}$.
Hardware upgrades to e.g. improve signal coupling onto the trap electrodes should enable sensing up to $\approx 400 \textrm{ MHz}$.

\subsection{Appendix E: The Minimum Frequency Uncertainty of an 80~MHz Signal Tone}

\begin{figure}[h!]
    \centering
    \includegraphics[width = 0.8\textwidth]{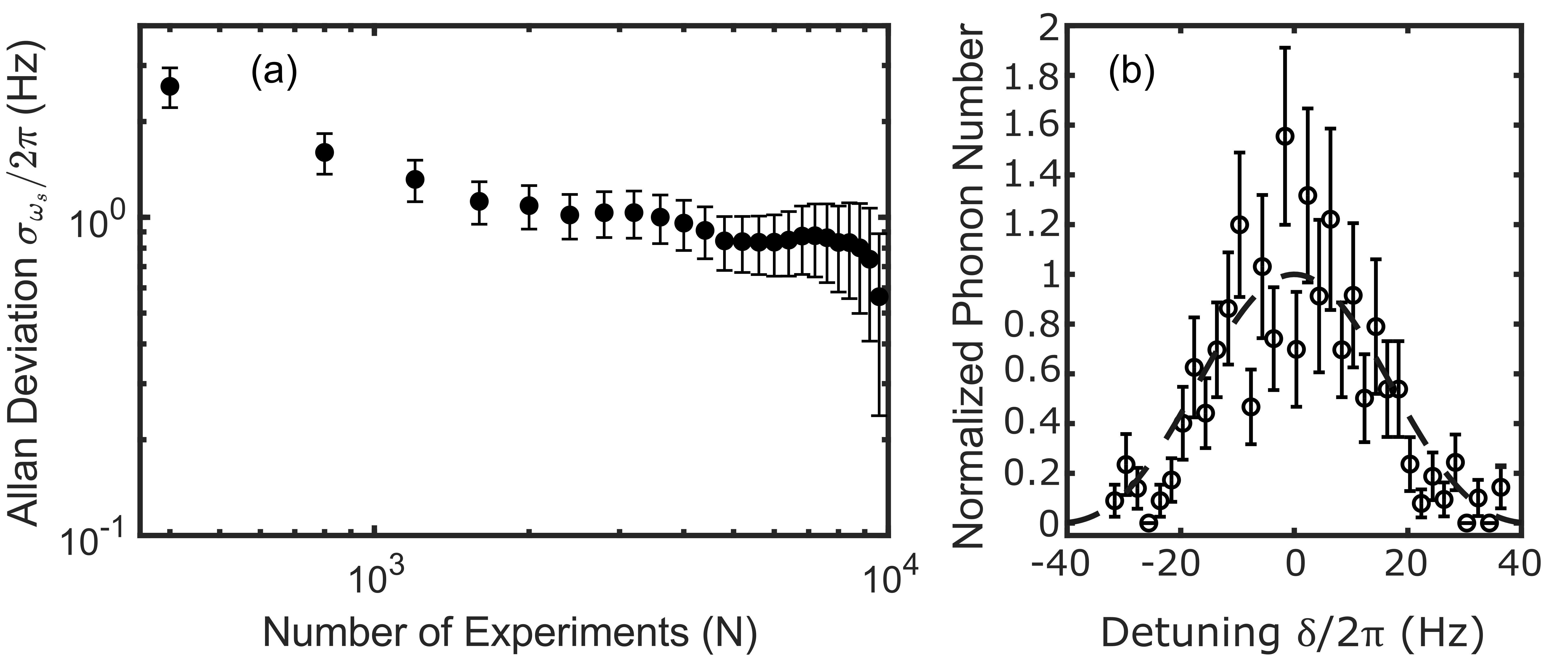}
    \caption{
    High resolution scan of an $80 \textrm{ MHz}$ probe tone using the $K=12$ subharmonic resonance on the axial mode.
    (a) An overlapped Allan deviation yields a minimum frequency uncertainty $\sigma_{\omega_s}/2\pi=0.56(32) \textrm{ Hz}$ at $N=9600$.
    The interrogation time is fixed at $\tau = 5 \textrm{ ms}$.
    Each individual experiment takes $22 \textrm{ ms}$ and is comprised of cooling pulses, the subharmonic excitation sequence, and state detection.
    (b) Motional excitation spectrum generated by scanning the frequency of the probe tone.
    Error bars represent one standard error.
    }
    \label{fig:min}
\end{figure}

To achieve a state of art frequency resolution for an $80 \textrm{ MHz}$ signal tone, we excited the $K=12$ subharmonic resonance onthe axial mode with an interrogation time up to $\tau = 5 \textrm{ ms}$, as illustrated in the inset of Fig.~\ref{fig:min}.
The Allan deviation was determined by measuring two points at the FHWM of the motional spectrum.
A minimum frequency resolution of $\sigma_{\omega_s}/2\pi=0.56(32) \textrm{ Hz}$ after $N=9600$ experiments.
The motional coherence time for the axial mode is $20 \textrm{ ms}$.
The primary limitation on the achievable minimum frequency resolution is the damage threshold of trap electronics(5~Watts).
Measurements at other values of 
$K$ at $\tau=5$~ms were not pursued to avoid risking damage to the electronic hardware.
Upgrades to system electronics would allow excitation of subharmonic orders beyond $K=12$, improving the minimum resolution by a factor of at least $> 4\times$.

\subsection{Appendix F: Sensitivity of Phonon Readout}

A common method for determining the mean phonon number $\langle n \rangle$ of different quantum states involves measuring the excited state population ($P_r$) after a pulse applied for $t=\pi/\Omega_{01}$ the red motional sideband of the $\ket{^2 S_{\frac{1}{2}}, m_J=-\frac{1}{2}} \Rightarrow \ket{^2 D_{\frac{5}{2}}, m_J=-\frac{5}{2}}$ qubit transition, where $\Omega_{01}$ is the Rabi frequency on the blue motional sideband on the same transition between Fock state n=0 and n=1.

\begin{figure}[h!]
    \centering
    \includegraphics[width = 0.8\textwidth]{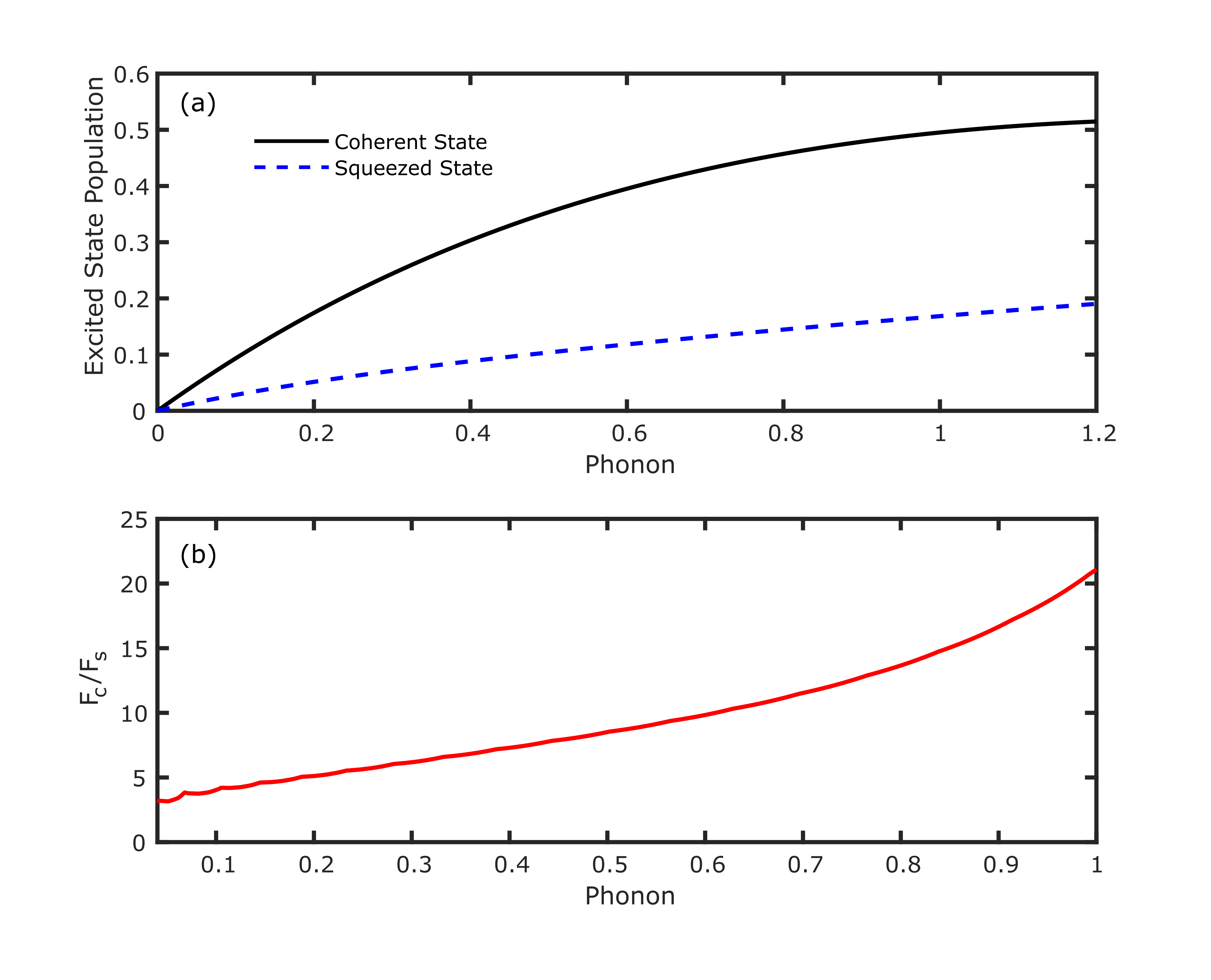}
    \caption{
    Characterizing the sensitivity of phonon readout for coherent and squeezed states.
    (a) The relation between population probability in excited state $\ket{^2 D_{\frac{5}{2}}}$ and phonon number for different motional states.
        The population is measured after applying red sideband Rabi oscillation for interrogation time $\tau=\pi/\Omega_{01}$, where $\Omega_{01}$ is the Rabi frequency for the $\ket{^2 S_{\frac{1}{2}},m_J=-\frac{1}{2},n=0}$  to $\ket{^2 D_{\frac{5}{2}},m_J=-\frac{5}{2},n=1}$ transition.
    (b) The Fisher information for a coherent state~($F_c$) is greater than the Fisher information of a squeezed state~($F_s$) for any mean phonon number, making them a more desirable quantum state for precision measurements.
    }
    \label{fig:detection}
\end{figure}

In Fig.~\ref{fig:detection}~(a), we plot the mapping of phonon number onto the excited state population for both a coherent state and a squeezed state using our protocol. 
As can be seen, due to the form of a squeezed state in the Fock basis, the slope of the mapping is much less steep for a squeezed state than a coherent state.
Intuitively, this means that squeezed states will provide worse performance during the readout step.
More quantitatively, in the framework of quantum metrology, better sensitivity corresponds to a larger Fisher information.
For the same mean phonon number, the Fisher information for readout of a squeezed state~($F_s$) is lower than for that of a coherent state~($F_c$) for our readout scheme.
Specifically, the Fisher information of our readout is $F{\left( n\right)}= \sum \frac{1}{P_r} \left(\frac{\partial P_r}{\partial n}\right)^2 = \frac{1}{P_r(1-P_r)}\left(\frac{\partial P_r}{\partial n}\right)^2$.
From the relation between excited state population ($P_r$) and phonon number shown in Fig.~\ref{fig:detection}(a), the ratio of the Fisher information between a coherent and a squeezed state is shown in Fig.~\ref{fig:detection}(b); the improved sensitivity of the coherent state is apparent as $F_{c}/F_{s}$ exceeds unity for all $\langle n \rangle$, demonstrating that coherent states yield higher Fisher information and therefore improved detection sensitivity, making them advantageous for precision measurement.

\subsection{Appendix G: Frequency Sensitivity under Dephasing with Different Strategies}

Though nonclassical states are used to improve sensitivities \cite{Burd2019, Wolf2019, McCormick2019, Gilmore2017, Hempel2013}, they incur a penalty of excess decoherence due to their sub-Planck structure \cite{Zurek2001, Turchette2000}.
This is well-noted with GHZ states \cite{RMP2017}, which can be used to enhance a signal by a factor of $M$, where $M$ is the number of probes in the GHZ state.
However, in the presence of pure dephasing, GHZ states decohere $M$ times faster, which negates their $M$-fold gain to yield \emph{no} net metrological gain.
Similarly, when sensing on bosonic modes (such as those of a trapped ion), Fock superposition states $\ket{\psi_{n}} = \frac{\ket{0} + \ket{n}}{\sqrt{2}}$ offer Heisenberg-limited $n$-fold gain, but decohere in excess by a factor $n^{2}$ for a net \emph{reduction} in metrological potential \cite{McCormick2019, Turchette2000, Myatt2000}.

In this context, our nonlinear protocol can be used to achieve \emph{ultimate} metrological gain as it yields amplification while still creating a coherent state - as coherent states are classical and lack the sub-Planck structure of nonclassical states, they incur no excess decoherence, which, when combined with our nonlinear protocol, allows for net, \emph{penalty-free} metrological gain.

To concretely compare the metrological potential of our nonlinear protocol in the presence of decoherence, we simulate the following motional Ramsey protocol for frequency sensing: the state is first displaced, then subject to free evolution for a delay time $T_{D}$, then antidisplaced \cite{Wolf2019, McCormick2019, McCormickDisplacedFock2019}.
The state is coupled to a phase reservoir over the entirety of the protocol to induce dephasing.
This coupling is selected to give $T_{2} = 2 \textrm{ ms}.$
Amplification strategies using nonclassical states typically apply this motional Ramsey on the state, then apply time-reversal dynamics (i.e. undo the nonclassical state) to remove any added noise \cite{Burd2019, Burd2021}.
For example, quantum-enhanced frequency sensing with fock states uses a protocol of state creation ($\ket{0} \rightarrow \ket{n}$), displacement, free evolution, antidisplacement, and state reversal ($\ket{n} \rightarrow \ket{0}$) \cite{Wolf2019}.
For simplicity, we assume the nonclassical state creation and reversal to be instantaneous.
By contrast, our nonlinear subharmonic protocol involves no nonclassical state dynamics; instead, the displacement (and antidisplacement) is simply generated using a higher-order subharmonic resonance - here, we use the $K=10$ subharmonic..
As a reference, we show the bare $K=2$ linear protocol, which represents the standard quantum limit.
For fair comparison between nonclassical protocols and our nonlinear protocol, the displacement of the motional Ramsey is kept constant at $\alpha=0.1$, while the amplification of the nonclassical states (characterized by the mean phonon number $\overline{n}$ of the amplified state) is kept constant across the nonclassical states, i.e. $\overline{n}=5$.

To determine the ultimate achievable metrological gain of each protocol, we simulate a frequency linescan at each delay time $T_{D}$.
The maximum frequency Fisher information at each delay time is then calculated from the linescan using $F{\left(\delta\right)} = \frac{\left( \partial_{\delta}{P} \right)^{2}}{\textrm{var}{\left( P \right)}}$ \cite{Wolf2019}, from which the sensitivity $\sigma_{\delta}$ can be extracted (via $\sigma_{\delta} = \sqrt{\frac{1}{F{\left(\delta\right)}/T}}$).

\begin{figure*}[h!]
    \centering
    \includegraphics[width=0.8\textwidth]{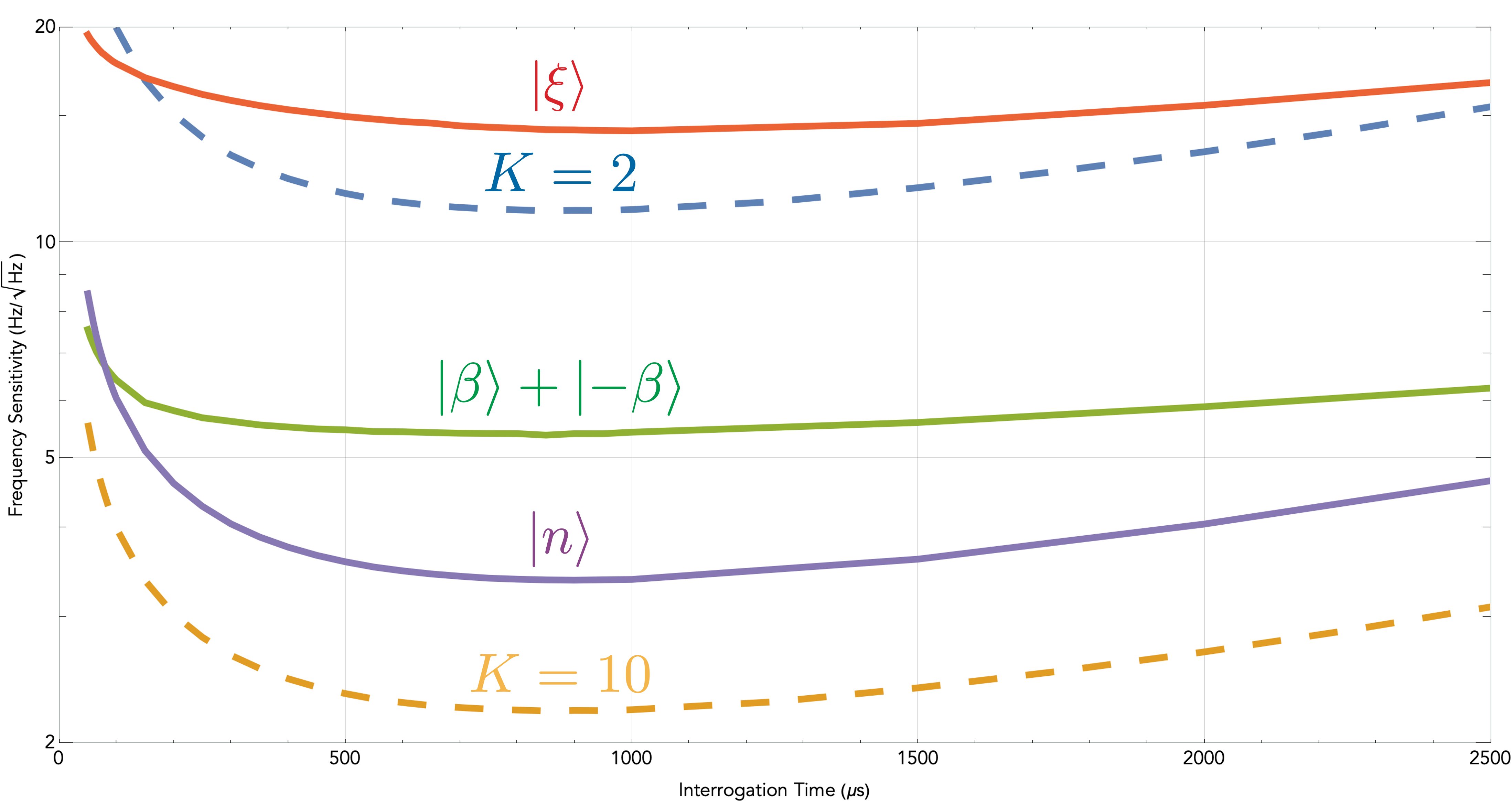}
    \caption{
    Simulated frequency sensitivity using a motional Ramsey protocol (displace - delay - antidisplace) with different amplification strategies in the presence of dephasing ($T_{2} = 2 \textrm{ ms}$).
    The amplification-free case is represented by the $\ket{K=2}$ linear protocol, which uses $\ket{\alpha=0.1}$.
    Quantum amplification with nonclassical states is implemented by applying this $K=2$, $\ket{\alpha=0.1}$ linear protocol to the nonclassical state, then time-reversing the nonclassical dynamics.
    We simulate quantum amplification of the motional Ramsey using a Fock state ($\ket{n}$, $n=5$), a squeezed state ($\ket{\xi}$, $\xi=1.544$), and a cat state ($\ket{\beta} + \ket{-\beta}$, $\beta=2.236$).
    The parameters of the nonclassical states have been chosen to keep the mean phonon number $\overline{n}=5$.
    Amplification with our nonlinear protocol is implemented by instead generating the displacement of the motional Ramsey using a higher-order subharmonic, shown here as the $K=10$ subharmonic.
    For fair comparison, the displacement is kept the same as that with the nonclassical protocols, i.e. $\alpha=0.1$.
    }
    \label{fig:decoherence_gain_comparison}
\end{figure*}

From Fig.~\ref{fig:decoherence_gain_comparison}, though Fock states achieve the best sensitivity when using a nonclassical strategy, this improvement is surpassed by our nonlinear protocol using the $K=5$ subharmonic.
Again, this is due to the fact that our nonlinear protocol results in the creation of a classical coherent state, which suffers no excess decoherence.
In this sense, such a nonlinear protocol can be said to be \emph{penalty-free}.

\begin{thebibliography}{47}%
\makeatletter
\providecommand \@ifxundefined [1]{%
 \@ifx{#1\undefined}
}%
\providecommand \@ifnum [1]{%
 \ifnum #1\expandafter \@firstoftwo
 \else \expandafter \@secondoftwo
 \fi
}%
\providecommand \@ifx [1]{%
 \ifx #1\expandafter \@firstoftwo
 \else \expandafter \@secondoftwo
 \fi
}%
\providecommand \natexlab [1]{#1}%
\providecommand \enquote  [1]{``#1''}%
\providecommand \bibnamefont  [1]{#1}%
\providecommand \bibfnamefont [1]{#1}%
\providecommand \citenamefont [1]{#1}%
\providecommand \href@noop [0]{\@secondoftwo}%
\providecommand \href [0]{\begingroup \@sanitize@url \@href}%
\providecommand \@href[1]{\@@startlink{#1}\@@href}%
\providecommand \@@href[1]{\endgroup#1\@@endlink}%
\providecommand \@sanitize@url [0]{\catcode `\\12\catcode `\$12\catcode `\&12\catcode `\#12\catcode `\^12\catcode `\_12\catcode `\%12\relax}%
\providecommand \@@startlink[1]{}%
\providecommand \@@endlink[0]{}%
\providecommand \url  [0]{\begingroup\@sanitize@url \@url }%
\providecommand \@url [1]{\endgroup\@href {#1}{\urlprefix }}%
\providecommand \urlprefix  [0]{URL }%
\providecommand \Eprint [0]{\href }%
\providecommand \doibase [0]{https://doi.org/}%
\providecommand \selectlanguage [0]{\@gobble}%
\providecommand \bibinfo  [0]{\@secondoftwo}%
\providecommand \bibfield  [0]{\@secondoftwo}%
\providecommand \translation [1]{[#1]}%
\providecommand \BibitemOpen [0]{}%
\providecommand \bibitemStop [0]{}%
\providecommand \bibitemNoStop [0]{.\EOS\space}%
\providecommand \EOS [0]{\spacefactor3000\relax}%
\providecommand \BibitemShut  [1]{\csname bibitem#1\endcsname}%
\let\auto@bib@innerbib\@empty
\bibitem [{\citenamefont {Hempel}\ \emph {et~al.}(2013)\citenamefont {Hempel}, \citenamefont {Lanyon}, \citenamefont {Jurcevic}, \citenamefont {Gerritsma}, \citenamefont {Blatt},\ and\ \citenamefont {Roos}}]{Hempel2013}%
  \BibitemOpen
  \bibfield  {author} {\bibinfo {author} {\bibfnamefont {C.}~\bibnamefont {Hempel}}, \bibinfo {author} {\bibfnamefont {B.~P.}\ \bibnamefont {Lanyon}}, \bibinfo {author} {\bibfnamefont {P.}~\bibnamefont {Jurcevic}}, \bibinfo {author} {\bibfnamefont {R.}~\bibnamefont {Gerritsma}}, \bibinfo {author} {\bibfnamefont {R.}~\bibnamefont {Blatt}},\ and\ \bibinfo {author} {\bibfnamefont {C.~F.}\ \bibnamefont {Roos}},\ }\href {https://doi.org/10.1038/nphoton.2013.172} {\bibfield  {journal} {\bibinfo  {journal} {Nature Photonics}\ }\textbf {\bibinfo {volume} {7}},\ \bibinfo {pages} {630} (\bibinfo {year} {2013})}\BibitemShut {NoStop}%
\bibitem [{\citenamefont {Burd}\ \emph {et~al.}(2019)\citenamefont {Burd}, \citenamefont {Srinivas}, \citenamefont {Bollinger}, \citenamefont {Wilson}, \citenamefont {Wineland}, \citenamefont {Leibfried}, \citenamefont {Slichter},\ and\ \citenamefont {Allcock}}]{Burd2019}%
  \BibitemOpen
  \bibfield  {author} {\bibinfo {author} {\bibfnamefont {S.~C.}\ \bibnamefont {Burd}}, \bibinfo {author} {\bibfnamefont {R.}~\bibnamefont {Srinivas}}, \bibinfo {author} {\bibfnamefont {J.~J.}\ \bibnamefont {Bollinger}}, \bibinfo {author} {\bibfnamefont {A.~C.}\ \bibnamefont {Wilson}}, \bibinfo {author} {\bibfnamefont {D.~J.}\ \bibnamefont {Wineland}}, \bibinfo {author} {\bibfnamefont {D.}~\bibnamefont {Leibfried}}, \bibinfo {author} {\bibfnamefont {D.~H.}\ \bibnamefont {Slichter}},\ and\ \bibinfo {author} {\bibfnamefont {D.~T.~C.}\ \bibnamefont {Allcock}},\ }\href {https://doi.org/10.1126/science.aaw2884} {\bibfield  {journal} {\bibinfo  {journal} {Science}\ }\textbf {\bibinfo {volume} {364}},\ \bibinfo {pages} {1163} (\bibinfo {year} {2019})},\ \bibinfo {note} {doi: 10.1126/science.aaw2884}\BibitemShut {NoStop}%
\bibitem [{\citenamefont {Wolf}\ \emph {et~al.}(2019)\citenamefont {Wolf}, \citenamefont {Shi}, \citenamefont {Heip}, \citenamefont {Gessner}, \citenamefont {Pezzè}, \citenamefont {Smerzi}, \citenamefont {Schulte}, \citenamefont {Hammerer},\ and\ \citenamefont {Schmidt}}]{Wolf2019}%
  \BibitemOpen
  \bibfield  {author} {\bibinfo {author} {\bibfnamefont {F.}~\bibnamefont {Wolf}}, \bibinfo {author} {\bibfnamefont {C.}~\bibnamefont {Shi}}, \bibinfo {author} {\bibfnamefont {J.~C.}\ \bibnamefont {Heip}}, \bibinfo {author} {\bibfnamefont {M.}~\bibnamefont {Gessner}}, \bibinfo {author} {\bibfnamefont {L.}~\bibnamefont {Pezzè}}, \bibinfo {author} {\bibfnamefont {A.}~\bibnamefont {Smerzi}}, \bibinfo {author} {\bibfnamefont {M.}~\bibnamefont {Schulte}}, \bibinfo {author} {\bibfnamefont {K.}~\bibnamefont {Hammerer}},\ and\ \bibinfo {author} {\bibfnamefont {P.~O.}\ \bibnamefont {Schmidt}},\ }\href {https://doi.org/10.1038/s41467-019-10576-4} {\bibfield  {journal} {\bibinfo  {journal} {Nature Communications}\ }\textbf {\bibinfo {volume} {10}},\ \bibinfo {pages} {2929} (\bibinfo {year} {2019})}\BibitemShut {NoStop}%
\bibitem [{\citenamefont {Kn\"unz}\ \emph {et~al.}(2010)\citenamefont {Kn\"unz}, \citenamefont {Herrmann}, \citenamefont {Batteiger}, \citenamefont {Saathoff}, \citenamefont {H\"ansch}, \citenamefont {Vahala},\ and\ \citenamefont {Udem}}]{Knunz2010}%
  \BibitemOpen
  \bibfield  {author} {\bibinfo {author} {\bibfnamefont {S.}~\bibnamefont {Kn\"unz}}, \bibinfo {author} {\bibfnamefont {M.}~\bibnamefont {Herrmann}}, \bibinfo {author} {\bibfnamefont {V.}~\bibnamefont {Batteiger}}, \bibinfo {author} {\bibfnamefont {G.}~\bibnamefont {Saathoff}}, \bibinfo {author} {\bibfnamefont {T.~W.}\ \bibnamefont {H\"ansch}}, \bibinfo {author} {\bibfnamefont {K.}~\bibnamefont {Vahala}},\ and\ \bibinfo {author} {\bibfnamefont {T.}~\bibnamefont {Udem}},\ }\href {https://doi.org/10.1103/PhysRevLett.105.013004} {\bibfield  {journal} {\bibinfo  {journal} {Phys. Rev. Lett.}\ }\textbf {\bibinfo {volume} {105}},\ \bibinfo {pages} {013004} (\bibinfo {year} {2010})}\BibitemShut {NoStop}%
\bibitem [{\citenamefont {Biercuk}\ \emph {et~al.}(2010)\citenamefont {Biercuk}, \citenamefont {Uys}, \citenamefont {Britton}, \citenamefont {VanDevender},\ and\ \citenamefont {Bollinger}}]{Biercuk2010}%
  \BibitemOpen
  \bibfield  {author} {\bibinfo {author} {\bibfnamefont {M.~J.}\ \bibnamefont {Biercuk}}, \bibinfo {author} {\bibfnamefont {H.}~\bibnamefont {Uys}}, \bibinfo {author} {\bibfnamefont {J.~W.}\ \bibnamefont {Britton}}, \bibinfo {author} {\bibfnamefont {A.~P.}\ \bibnamefont {VanDevender}},\ and\ \bibinfo {author} {\bibfnamefont {J.~J.}\ \bibnamefont {Bollinger}},\ }\href {https://doi.org/10.1038/nnano.2010.165} {\bibfield  {journal} {\bibinfo  {journal} {Nature Nanotechnology}\ }\textbf {\bibinfo {volume} {5}},\ \bibinfo {pages} {646} (\bibinfo {year} {2010})}\BibitemShut {NoStop}%
\bibitem [{\citenamefont {Gilmore}\ \emph {et~al.}(2017)\citenamefont {Gilmore}, \citenamefont {Bohnet}, \citenamefont {Sawyer}, \citenamefont {Britton},\ and\ \citenamefont {Bollinger}}]{Gilmore2017}%
  \BibitemOpen
  \bibfield  {author} {\bibinfo {author} {\bibfnamefont {K.~A.}\ \bibnamefont {Gilmore}}, \bibinfo {author} {\bibfnamefont {J.~G.}\ \bibnamefont {Bohnet}}, \bibinfo {author} {\bibfnamefont {B.~C.}\ \bibnamefont {Sawyer}}, \bibinfo {author} {\bibfnamefont {J.~W.}\ \bibnamefont {Britton}},\ and\ \bibinfo {author} {\bibfnamefont {J.~J.}\ \bibnamefont {Bollinger}},\ }\href {https://doi.org/10.1103/PhysRevLett.118.263602} {\bibfield  {journal} {\bibinfo  {journal} {Phys. Rev. Lett.}\ }\textbf {\bibinfo {volume} {118}},\ \bibinfo {pages} {263602} (\bibinfo {year} {2017})}\BibitemShut {NoStop}%
\bibitem [{\citenamefont {Liu}\ \emph {et~al.}(2021)\citenamefont {Liu}, \citenamefont {Wei}, \citenamefont {Chen}, \citenamefont {Li}, \citenamefont {Dai}, \citenamefont {Zhou},\ and\ \citenamefont {Feng}}]{Liu2021}%
  \BibitemOpen
  \bibfield  {author} {\bibinfo {author} {\bibfnamefont {Z.}~\bibnamefont {Liu}}, \bibinfo {author} {\bibfnamefont {Y.}~\bibnamefont {Wei}}, \bibinfo {author} {\bibfnamefont {L.}~\bibnamefont {Chen}}, \bibinfo {author} {\bibfnamefont {J.}~\bibnamefont {Li}}, \bibinfo {author} {\bibfnamefont {S.}~\bibnamefont {Dai}}, \bibinfo {author} {\bibfnamefont {F.}~\bibnamefont {Zhou}},\ and\ \bibinfo {author} {\bibfnamefont {M.}~\bibnamefont {Feng}},\ }\href {https://doi.org/10.1103/PhysRevApplied.16.044007} {\bibfield  {journal} {\bibinfo  {journal} {Phys. Rev. Appl.}\ }\textbf {\bibinfo {volume} {16}},\ \bibinfo {pages} {044007} (\bibinfo {year} {2021})}\BibitemShut {NoStop}%
\bibitem [{\citenamefont {Campbell}\ and\ \citenamefont {Hamilton}(2017)}]{Campbell2017}%
  \BibitemOpen
  \bibfield  {author} {\bibinfo {author} {\bibfnamefont {W.~C.}\ \bibnamefont {Campbell}}\ and\ \bibinfo {author} {\bibfnamefont {P.}~\bibnamefont {Hamilton}},\ }\href {https://doi.org/10.1088/1361-6455/aa5a8f} {\bibfield  {journal} {\bibinfo  {journal} {Journal of Physics B: Atomic, Molecular and Optical Physics}\ }\textbf {\bibinfo {volume} {50}},\ \bibinfo {pages} {064002} (\bibinfo {year} {2017})}\BibitemShut {NoStop}%
\bibitem [{\citenamefont {Bradley}\ \emph {et~al.}(2003)\citenamefont {Bradley}, \citenamefont {Clarke}, \citenamefont {Kinion}, \citenamefont {Rosenberg}, \citenamefont {van Bibber}, \citenamefont {Matsuki}, \citenamefont {M\"uck},\ and\ \citenamefont {Sikivie}}]{Bradley2003}%
  \BibitemOpen
  \bibfield  {author} {\bibinfo {author} {\bibfnamefont {R.}~\bibnamefont {Bradley}}, \bibinfo {author} {\bibfnamefont {J.}~\bibnamefont {Clarke}}, \bibinfo {author} {\bibfnamefont {D.}~\bibnamefont {Kinion}}, \bibinfo {author} {\bibfnamefont {L.~J.}\ \bibnamefont {Rosenberg}}, \bibinfo {author} {\bibfnamefont {K.}~\bibnamefont {van Bibber}}, \bibinfo {author} {\bibfnamefont {S.}~\bibnamefont {Matsuki}}, \bibinfo {author} {\bibfnamefont {M.}~\bibnamefont {M\"uck}},\ and\ \bibinfo {author} {\bibfnamefont {P.}~\bibnamefont {Sikivie}},\ }\href {https://doi.org/10.1103/RevModPhys.75.777} {\bibfield  {journal} {\bibinfo  {journal} {Rev. Mod. Phys.}\ }\textbf {\bibinfo {volume} {75}},\ \bibinfo {pages} {777} (\bibinfo {year} {2003})}\BibitemShut {NoStop}%
\bibitem [{\citenamefont {Abbott}\ \emph {et~al.}(2016)\citenamefont {Abbott} \emph {et~al.}}]{LIGO}%
  \BibitemOpen
  \bibfield  {author} {\bibinfo {author} {\bibfnamefont {B.~P.}\ \bibnamefont {Abbott}} \emph {et~al.} (\bibinfo {collaboration} {LIGO Scientific Collaboration and Virgo Collaboration}),\ }\href {https://doi.org/10.1103/PhysRevLett.116.061102} {\bibfield  {journal} {\bibinfo  {journal} {Phys. Rev. Lett.}\ }\textbf {\bibinfo {volume} {116}},\ \bibinfo {pages} {061102} (\bibinfo {year} {2016})}\BibitemShut {NoStop}%
\bibitem [{\citenamefont {Wu}\ \emph {et~al.}(2025)\citenamefont {Wu}, \citenamefont {Mitts}, \citenamefont {Ho}, \citenamefont {Rabinowitz},\ and\ \citenamefont {Hudson}}]{Wu2025}%
  \BibitemOpen
  \bibfield  {author} {\bibinfo {author} {\bibfnamefont {H.}~\bibnamefont {Wu}}, \bibinfo {author} {\bibfnamefont {G.~D.}\ \bibnamefont {Mitts}}, \bibinfo {author} {\bibfnamefont {C.~Z.~C.}\ \bibnamefont {Ho}}, \bibinfo {author} {\bibfnamefont {J.~A.}\ \bibnamefont {Rabinowitz}},\ and\ \bibinfo {author} {\bibfnamefont {E.~R.}\ \bibnamefont {Hudson}},\ }\href {https://doi.org/10.1038/s41567-024-02753-0} {\bibfield  {journal} {\bibinfo  {journal} {Nature Physics}\ }\textbf {\bibinfo {volume} {21}},\ \bibinfo {pages} {380} (\bibinfo {year} {2025})}\BibitemShut {NoStop}%
\bibitem [{\citenamefont {Huelga}\ \emph {et~al.}(1997)\citenamefont {Huelga}, \citenamefont {Macchiavello}, \citenamefont {Pellizzari}, \citenamefont {Ekert}, \citenamefont {Plenio},\ and\ \citenamefont {Cirac}}]{Huelga1997}%
  \BibitemOpen
  \bibfield  {author} {\bibinfo {author} {\bibfnamefont {S.~F.}\ \bibnamefont {Huelga}}, \bibinfo {author} {\bibfnamefont {C.}~\bibnamefont {Macchiavello}}, \bibinfo {author} {\bibfnamefont {T.}~\bibnamefont {Pellizzari}}, \bibinfo {author} {\bibfnamefont {A.~K.}\ \bibnamefont {Ekert}}, \bibinfo {author} {\bibfnamefont {M.~B.}\ \bibnamefont {Plenio}},\ and\ \bibinfo {author} {\bibfnamefont {J.~I.}\ \bibnamefont {Cirac}},\ }\href {https://doi.org/10.1103/PhysRevLett.79.3865} {\bibfield  {journal} {\bibinfo  {journal} {Phys. Rev. Lett.}\ }\textbf {\bibinfo {volume} {79}},\ \bibinfo {pages} {3865} (\bibinfo {year} {1997})}\BibitemShut {NoStop}%
\bibitem [{\citenamefont {Johnson}\ \emph {et~al.}(2017)\citenamefont {Johnson}, \citenamefont {Wong-Campos}, \citenamefont {Neyenhuis}, \citenamefont {Mizrahi},\ and\ \citenamefont {Monroe}}]{Johnson2017}%
  \BibitemOpen
  \bibfield  {author} {\bibinfo {author} {\bibfnamefont {K.~G.}\ \bibnamefont {Johnson}}, \bibinfo {author} {\bibfnamefont {J.~D.}\ \bibnamefont {Wong-Campos}}, \bibinfo {author} {\bibfnamefont {B.}~\bibnamefont {Neyenhuis}}, \bibinfo {author} {\bibfnamefont {J.}~\bibnamefont {Mizrahi}},\ and\ \bibinfo {author} {\bibfnamefont {C.}~\bibnamefont {Monroe}},\ }\href {https://doi.org/10.1038/s41467-017-00682-6} {\bibfield  {journal} {\bibinfo  {journal} {Nature Communications}\ }\textbf {\bibinfo {volume} {8}},\ \bibinfo {pages} {697} (\bibinfo {year} {2017})}\BibitemShut {NoStop}%
\bibitem [{\citenamefont {Monroe}\ \emph {et~al.}(1996)\citenamefont {Monroe}, \citenamefont {Meekhof}, \citenamefont {King},\ and\ \citenamefont {Wineland}}]{Monroe1996}%
  \BibitemOpen
  \bibfield  {author} {\bibinfo {author} {\bibfnamefont {C.}~\bibnamefont {Monroe}}, \bibinfo {author} {\bibfnamefont {D.~M.}\ \bibnamefont {Meekhof}}, \bibinfo {author} {\bibfnamefont {B.~E.}\ \bibnamefont {King}},\ and\ \bibinfo {author} {\bibfnamefont {D.~J.}\ \bibnamefont {Wineland}},\ }\href {https://doi.org/10.1126/science.272.5265.1131} {\bibfield  {journal} {\bibinfo  {journal} {Science}\ }\textbf {\bibinfo {volume} {272}},\ \bibinfo {pages} {1131} (\bibinfo {year} {1996})}\BibitemShut {NoStop}%
\bibitem [{\citenamefont {McCormick}\ \emph {et~al.}(2019{\natexlab{a}})\citenamefont {McCormick}, \citenamefont {Keller}, \citenamefont {Burd}, \citenamefont {Wineland}, \citenamefont {Wilson},\ and\ \citenamefont {Leibfried}}]{McCormick2019}%
  \BibitemOpen
  \bibfield  {author} {\bibinfo {author} {\bibfnamefont {K.~C.}\ \bibnamefont {McCormick}}, \bibinfo {author} {\bibfnamefont {J.}~\bibnamefont {Keller}}, \bibinfo {author} {\bibfnamefont {S.~C.}\ \bibnamefont {Burd}}, \bibinfo {author} {\bibfnamefont {D.~J.}\ \bibnamefont {Wineland}}, \bibinfo {author} {\bibfnamefont {A.~C.}\ \bibnamefont {Wilson}},\ and\ \bibinfo {author} {\bibfnamefont {D.}~\bibnamefont {Leibfried}},\ }\href {https://doi.org/10.1038/s41586-019-1421-y} {\bibfield  {journal} {\bibinfo  {journal} {Nature}\ }\textbf {\bibinfo {volume} {572}},\ \bibinfo {pages} {86} (\bibinfo {year} {2019}{\natexlab{a}})}\BibitemShut {NoStop}%
\bibitem [{\citenamefont {Wolf}(2018)}]{Wolfthesis}%
  \BibitemOpen
  \bibfield  {author} {\bibinfo {author} {\bibfnamefont {F.}~\bibnamefont {Wolf}},\ }\emph {\bibinfo {title} {Motional quantum state engineering for quantum logic spectroscopy of molecular ions}},\ \href@noop {} {Ph.D. thesis},\ \bibinfo  {school} {Leibniz Universit at Hannover} (\bibinfo {year} {2018})\BibitemShut {NoStop}%
\bibitem [{\citenamefont {Pezz\`e}\ \emph {et~al.}(2018)\citenamefont {Pezz\`e}, \citenamefont {Smerzi}, \citenamefont {Oberthaler}, \citenamefont {Schmied},\ and\ \citenamefont {Treutlein}}]{Pezze2018}%
  \BibitemOpen
  \bibfield  {author} {\bibinfo {author} {\bibfnamefont {L.}~\bibnamefont {Pezz\`e}}, \bibinfo {author} {\bibfnamefont {A.}~\bibnamefont {Smerzi}}, \bibinfo {author} {\bibfnamefont {M.~K.}\ \bibnamefont {Oberthaler}}, \bibinfo {author} {\bibfnamefont {R.}~\bibnamefont {Schmied}},\ and\ \bibinfo {author} {\bibfnamefont {P.}~\bibnamefont {Treutlein}},\ }\href {https://doi.org/10.1103/RevModPhys.90.035005} {\bibfield  {journal} {\bibinfo  {journal} {Rev. Mod. Phys.}\ }\textbf {\bibinfo {volume} {90}},\ \bibinfo {pages} {035005} (\bibinfo {year} {2018})}\BibitemShut {NoStop}%
\bibitem [{\citenamefont {Boixo}\ \emph {et~al.}(2007)\citenamefont {Boixo}, \citenamefont {Flammia}, \citenamefont {Caves},\ and\ \citenamefont {Geremia}}]{Boixo2007}%
  \BibitemOpen
  \bibfield  {author} {\bibinfo {author} {\bibfnamefont {S.}~\bibnamefont {Boixo}}, \bibinfo {author} {\bibfnamefont {S.~T.}\ \bibnamefont {Flammia}}, \bibinfo {author} {\bibfnamefont {C.~M.}\ \bibnamefont {Caves}},\ and\ \bibinfo {author} {\bibfnamefont {J.}~\bibnamefont {Geremia}},\ }\href {https://doi.org/10.1103/PhysRevLett.98.090401} {\bibfield  {journal} {\bibinfo  {journal} {Phys. Rev. Lett.}\ }\textbf {\bibinfo {volume} {98}},\ \bibinfo {pages} {090401} (\bibinfo {year} {2007})}\BibitemShut {NoStop}%
\bibitem [{\citenamefont {Boixo}\ \emph {et~al.}(2008)\citenamefont {Boixo}, \citenamefont {Datta}, \citenamefont {Davis}, \citenamefont {Flammia}, \citenamefont {Shaji},\ and\ \citenamefont {Caves}}]{Boixo2008}%
  \BibitemOpen
  \bibfield  {author} {\bibinfo {author} {\bibfnamefont {S.}~\bibnamefont {Boixo}}, \bibinfo {author} {\bibfnamefont {A.}~\bibnamefont {Datta}}, \bibinfo {author} {\bibfnamefont {M.~J.}\ \bibnamefont {Davis}}, \bibinfo {author} {\bibfnamefont {S.~T.}\ \bibnamefont {Flammia}}, \bibinfo {author} {\bibfnamefont {A.}~\bibnamefont {Shaji}},\ and\ \bibinfo {author} {\bibfnamefont {C.~M.}\ \bibnamefont {Caves}},\ }\href {https://doi.org/10.1103/PhysRevLett.101.040403} {\bibfield  {journal} {\bibinfo  {journal} {Phys. Rev. Lett.}\ }\textbf {\bibinfo {volume} {101}},\ \bibinfo {pages} {040403} (\bibinfo {year} {2008})}\BibitemShut {NoStop}%
\bibitem [{\citenamefont {Boixo}\ \emph {et~al.}(2009)\citenamefont {Boixo}, \citenamefont {Datta}, \citenamefont {Davis}, \citenamefont {Shaji}, \citenamefont {Tacla},\ and\ \citenamefont {Caves}}]{Boixo2009}%
  \BibitemOpen
  \bibfield  {author} {\bibinfo {author} {\bibfnamefont {S.}~\bibnamefont {Boixo}}, \bibinfo {author} {\bibfnamefont {A.}~\bibnamefont {Datta}}, \bibinfo {author} {\bibfnamefont {M.~J.}\ \bibnamefont {Davis}}, \bibinfo {author} {\bibfnamefont {A.}~\bibnamefont {Shaji}}, \bibinfo {author} {\bibfnamefont {A.~B.}\ \bibnamefont {Tacla}},\ and\ \bibinfo {author} {\bibfnamefont {C.~M.}\ \bibnamefont {Caves}},\ }\href {https://doi.org/10.1103/PhysRevA.80.032103} {\bibfield  {journal} {\bibinfo  {journal} {Phys. Rev. A}\ }\textbf {\bibinfo {volume} {80}},\ \bibinfo {pages} {032103} (\bibinfo {year} {2009})}\BibitemShut {NoStop}%
\bibitem [{\citenamefont {Huang}\ \emph {et~al.}(2024)\citenamefont {Huang}, \citenamefont {Zhuang},\ and\ \citenamefont {Lee}}]{Huang2024}%
  \BibitemOpen
  \bibfield  {author} {\bibinfo {author} {\bibfnamefont {J.}~\bibnamefont {Huang}}, \bibinfo {author} {\bibfnamefont {M.}~\bibnamefont {Zhuang}},\ and\ \bibinfo {author} {\bibfnamefont {C.}~\bibnamefont {Lee}},\ }\href {https://doi.org/10.1063/5.0204102} {\bibfield  {journal} {\bibinfo  {journal} {Applied Physics Reviews}\ }\textbf {\bibinfo {volume} {11}},\ \bibinfo {pages} {031302} (\bibinfo {year} {2024})}\BibitemShut {NoStop}%
\bibitem [{SI()}]{SI}%
  \BibitemOpen
  \href@noop {} {}\bibinfo {note} {See Supplemental Material at [url] for details on theoretical analysis, parameter estimation, wideband demonstration, sensitivity of phonon readout, and frequency sensitivity under dephasing with different strategies, which includes Refs.~[23--29].}\BibitemShut {Stop}%
\bibitem [{\citenamefont {Johansson}\ \emph {et~al.}(2012)\citenamefont {Johansson}, \citenamefont {Nation},\ and\ \citenamefont {Nori}}]{JOHANSSON2012}%
  \BibitemOpen
  \bibfield  {author} {\bibinfo {author} {\bibfnamefont {J.}~\bibnamefont {Johansson}}, \bibinfo {author} {\bibfnamefont {P.}~\bibnamefont {Nation}},\ and\ \bibinfo {author} {\bibfnamefont {F.}~\bibnamefont {Nori}},\ }\href {https://doi.org/https://doi.org/10.1016/j.cpc.2012.02.021} {\bibfield  {journal} {\bibinfo  {journal} {Computer Physics Communications}\ }\textbf {\bibinfo {volume} {183}},\ \bibinfo {pages} {1760} (\bibinfo {year} {2012})}\BibitemShut {NoStop}%
\bibitem [{\citenamefont {Johansson}\ \emph {et~al.}(2013)\citenamefont {Johansson}, \citenamefont {Nation},\ and\ \citenamefont {Nori}}]{JOHANSSON2013}%
  \BibitemOpen
  \bibfield  {author} {\bibinfo {author} {\bibfnamefont {J.}~\bibnamefont {Johansson}}, \bibinfo {author} {\bibfnamefont {P.}~\bibnamefont {Nation}},\ and\ \bibinfo {author} {\bibfnamefont {F.}~\bibnamefont {Nori}},\ }\href {https://doi.org/https://doi.org/10.1016/j.cpc.2012.11.019} {\bibfield  {journal} {\bibinfo  {journal} {Computer Physics Communications}\ }\textbf {\bibinfo {volume} {184}},\ \bibinfo {pages} {1234} (\bibinfo {year} {2013})}\BibitemShut {NoStop}%
\bibitem [{\citenamefont {Lee}\ \emph {et~al.}(2005)\citenamefont {Lee}, \citenamefont {Brickman}, \citenamefont {Deslauriers}, \citenamefont {Haljan}, \citenamefont {Duan},\ and\ \citenamefont {Monroe}}]{Lee2005}%
  \BibitemOpen
  \bibfield  {author} {\bibinfo {author} {\bibfnamefont {P.~J.}\ \bibnamefont {Lee}}, \bibinfo {author} {\bibfnamefont {K.-A.}\ \bibnamefont {Brickman}}, \bibinfo {author} {\bibfnamefont {L.}~\bibnamefont {Deslauriers}}, \bibinfo {author} {\bibfnamefont {P.~C.}\ \bibnamefont {Haljan}}, \bibinfo {author} {\bibfnamefont {L.-M.}\ \bibnamefont {Duan}},\ and\ \bibinfo {author} {\bibfnamefont {C.}~\bibnamefont {Monroe}},\ }\href {https://doi.org/10.1088/1464-4266/7/10/025} {\bibfield  {journal} {\bibinfo  {journal} {Journal of Optics B: Quantum and Semiclassical Optics}\ }\textbf {\bibinfo {volume} {7}},\ \bibinfo {pages} {S371} (\bibinfo {year} {2005})}\BibitemShut {NoStop}%
\bibitem [{\citenamefont {Zurek}(2001)}]{Zurek2001}%
  \BibitemOpen
  \bibfield  {author} {\bibinfo {author} {\bibfnamefont {W.~H.}\ \bibnamefont {Zurek}},\ }\href {https://doi.org/10.1038/35089017} {\bibfield  {journal} {\bibinfo  {journal} {Nature}\ }\textbf {\bibinfo {volume} {412}},\ \bibinfo {pages} {712} (\bibinfo {year} {2001})}\BibitemShut {NoStop}%
\bibitem [{\citenamefont {Degen}\ \emph {et~al.}(2017)\citenamefont {Degen}, \citenamefont {Reinhard},\ and\ \citenamefont {Cappellaro}}]{Degen2017}%
  \BibitemOpen
  \bibfield  {author} {\bibinfo {author} {\bibfnamefont {C.~L.}\ \bibnamefont {Degen}}, \bibinfo {author} {\bibfnamefont {F.}~\bibnamefont {Reinhard}},\ and\ \bibinfo {author} {\bibfnamefont {P.}~\bibnamefont {Cappellaro}},\ }\href {https://doi.org/10.1103/RevModPhys.89.035002} {\bibfield  {journal} {\bibinfo  {journal} {Rev. Mod. Phys.}\ }\textbf {\bibinfo {volume} {89}},\ \bibinfo {pages} {035002} (\bibinfo {year} {2017})}\BibitemShut {NoStop}%
\bibitem [{\citenamefont {McCormick}\ \emph {et~al.}(2019{\natexlab{b}})\citenamefont {McCormick}, \citenamefont {Keller}, \citenamefont {Wineland}, \citenamefont {Wilson},\ and\ \citenamefont {Leibfried}}]{McCormickDisplacedFock2019}%
  \BibitemOpen
  \bibfield  {author} {\bibinfo {author} {\bibfnamefont {K.~C.}\ \bibnamefont {McCormick}}, \bibinfo {author} {\bibfnamefont {J.}~\bibnamefont {Keller}}, \bibinfo {author} {\bibfnamefont {D.~J.}\ \bibnamefont {Wineland}}, \bibinfo {author} {\bibfnamefont {A.~C.}\ \bibnamefont {Wilson}},\ and\ \bibinfo {author} {\bibfnamefont {D.}~\bibnamefont {Leibfried}},\ }\href {https://doi.org/10.1088/2058-9565/ab0513} {\bibfield  {journal} {\bibinfo  {journal} {Quantum Science and Technology}\ }\textbf {\bibinfo {volume} {4}},\ \bibinfo {pages} {024010} (\bibinfo {year} {2019}{\natexlab{b}})}\BibitemShut {NoStop}%
\bibitem [{\citenamefont {Burd}\ \emph {et~al.}(2021)\citenamefont {Burd}, \citenamefont {Srinivas}, \citenamefont {Knaack}, \citenamefont {Ge}, \citenamefont {Wilson}, \citenamefont {Wineland}, \citenamefont {Leibfried}, \citenamefont {Bollinger}, \citenamefont {Allcock},\ and\ \citenamefont {Slichter}}]{Burd2021}%
  \BibitemOpen
  \bibfield  {author} {\bibinfo {author} {\bibfnamefont {S.~C.}\ \bibnamefont {Burd}}, \bibinfo {author} {\bibfnamefont {R.}~\bibnamefont {Srinivas}}, \bibinfo {author} {\bibfnamefont {H.~M.}\ \bibnamefont {Knaack}}, \bibinfo {author} {\bibfnamefont {W.}~\bibnamefont {Ge}}, \bibinfo {author} {\bibfnamefont {A.~C.}\ \bibnamefont {Wilson}}, \bibinfo {author} {\bibfnamefont {D.~J.}\ \bibnamefont {Wineland}}, \bibinfo {author} {\bibfnamefont {D.}~\bibnamefont {Leibfried}}, \bibinfo {author} {\bibfnamefont {J.~J.}\ \bibnamefont {Bollinger}}, \bibinfo {author} {\bibfnamefont {D.~T.~C.}\ \bibnamefont {Allcock}},\ and\ \bibinfo {author} {\bibfnamefont {D.~H.}\ \bibnamefont {Slichter}},\ }\href {https://doi.org/10.1038/s41567-021-01237-9} {\bibfield  {journal} {\bibinfo  {journal} {Nature Physics}\ }\textbf {\bibinfo {volume} {17}},\ \bibinfo {pages} {898} (\bibinfo {year} {2021})}\BibitemShut {NoStop}%
\bibitem [{\citenamefont {Landau}\ and\ \citenamefont {Lifshitz}(1976)}]{Landau1976Mechanics}%
  \BibitemOpen
  \bibfield  {author} {\bibinfo {author} {\bibfnamefont {L.~D.}\ \bibnamefont {Landau}}\ and\ \bibinfo {author} {\bibfnamefont {E.~M.}\ \bibnamefont {Lifshitz}},\ }\href {http://www.worldcat.org/isbn/0750628960} {\emph {\bibinfo {title} {Mechanics, Third Edition: Volume 1 (Course of Theoretical Physics)}}},\ \bibinfo {edition} {3rd}\ ed.\ (\bibinfo  {publisher} {Butterworth-Heinemann},\ \bibinfo {year} {1976})\BibitemShut {NoStop}%
\bibitem [{\citenamefont {Sudakov}\ \emph {et~al.}(2000)\citenamefont {Sudakov}, \citenamefont {Konenkov}, \citenamefont {Douglas},\ and\ \citenamefont {Glebova}}]{SUDAKOV2000}%
  \BibitemOpen
  \bibfield  {author} {\bibinfo {author} {\bibfnamefont {M.}~\bibnamefont {Sudakov}}, \bibinfo {author} {\bibfnamefont {N.}~\bibnamefont {Konenkov}}, \bibinfo {author} {\bibfnamefont {D.}~\bibnamefont {Douglas}},\ and\ \bibinfo {author} {\bibfnamefont {T.}~\bibnamefont {Glebova}},\ }\href {https://doi.org/https://doi.org/10.1016/S1044-0305(99)00111-7} {\bibfield  {journal} {\bibinfo  {journal} {Journal of the American Society for Mass Spectrometry}\ }\textbf {\bibinfo {volume} {11}},\ \bibinfo {pages} {10} (\bibinfo {year} {2000})}\BibitemShut {NoStop}%
\bibitem [{\citenamefont {Zhao}\ \emph {et~al.}(2002)\citenamefont {Zhao}, \citenamefont {Ryjkov},\ and\ \citenamefont {Schuessler}}]{Zhao2002}%
  \BibitemOpen
  \bibfield  {author} {\bibinfo {author} {\bibfnamefont {X.}~\bibnamefont {Zhao}}, \bibinfo {author} {\bibfnamefont {V.~L.}\ \bibnamefont {Ryjkov}},\ and\ \bibinfo {author} {\bibfnamefont {H.~A.}\ \bibnamefont {Schuessler}},\ }\href {https://doi.org/10.1103/PhysRevA.66.063414} {\bibfield  {journal} {\bibinfo  {journal} {Phys. Rev. A}\ }\textbf {\bibinfo {volume} {66}},\ \bibinfo {pages} {063414} (\bibinfo {year} {2002})}\BibitemShut {NoStop}%
\bibitem [{\citenamefont {Collings}\ and\ \citenamefont {Douglas}(2000)}]{COLLINGS2000}%
  \BibitemOpen
  \bibfield  {author} {\bibinfo {author} {\bibfnamefont {B.}~\bibnamefont {Collings}}\ and\ \bibinfo {author} {\bibfnamefont {D.}~\bibnamefont {Douglas}},\ }\href {https://doi.org/https://doi.org/10.1016/S1044-0305(00)00171-9} {\bibfield  {journal} {\bibinfo  {journal} {Journal of the American Society for Mass Spectrometry}\ }\textbf {\bibinfo {volume} {11}},\ \bibinfo {pages} {1016} (\bibinfo {year} {2000})}\BibitemShut {NoStop}%
\bibitem [{\citenamefont {Tommaseo}\ \emph {et~al.}(2003)\citenamefont {Tommaseo}, \citenamefont {Paasche}, \citenamefont {Angelescu},\ and\ \citenamefont {Werth}}]{Tommaseo2003}%
  \BibitemOpen
  \bibfield  {author} {\bibinfo {author} {\bibfnamefont {G.}~\bibnamefont {Tommaseo}}, \bibinfo {author} {\bibfnamefont {P.}~\bibnamefont {Paasche}}, \bibinfo {author} {\bibfnamefont {C.}~\bibnamefont {Angelescu}},\ and\ \bibinfo {author} {\bibfnamefont {G.}~\bibnamefont {Werth}},\ }\href {https://doi.org/10.1140/epjd/e2003-00296-0} {\bibfield  {journal} {\bibinfo  {journal} {Eur. Phys. J. D}\ }\textbf {\bibinfo {volume} {28}},\ \bibinfo {pages} {39} (\bibinfo {year} {2003})}\BibitemShut {NoStop}%
\bibitem [{\citenamefont {Trebino}\ and\ \citenamefont {Rahn}(1987)}]{Trebino1987}%
  \BibitemOpen
  \bibfield  {author} {\bibinfo {author} {\bibfnamefont {R.}~\bibnamefont {Trebino}}\ and\ \bibinfo {author} {\bibfnamefont {L.~A.}\ \bibnamefont {Rahn}},\ }\href {https://doi.org/10.1364/OL.12.000912} {\bibfield  {journal} {\bibinfo  {journal} {Opt. Lett.}\ }\textbf {\bibinfo {volume} {12}},\ \bibinfo {pages} {912} (\bibinfo {year} {1987})}\BibitemShut {NoStop}%
\bibitem [{\citenamefont {Cataliotti}\ \emph {et~al.}(2001)\citenamefont {Cataliotti}, \citenamefont {Scheunemann}, \citenamefont {H\"ansch},\ and\ \citenamefont {Weitz}}]{Cataliotti2001}%
  \BibitemOpen
  \bibfield  {author} {\bibinfo {author} {\bibfnamefont {F.~S.}\ \bibnamefont {Cataliotti}}, \bibinfo {author} {\bibfnamefont {R.}~\bibnamefont {Scheunemann}}, \bibinfo {author} {\bibfnamefont {T.~W.}\ \bibnamefont {H\"ansch}},\ and\ \bibinfo {author} {\bibfnamefont {M.}~\bibnamefont {Weitz}},\ }\href {https://doi.org/10.1103/PhysRevLett.87.113601} {\bibfield  {journal} {\bibinfo  {journal} {Phys. Rev. Lett.}\ }\textbf {\bibinfo {volume} {87}},\ \bibinfo {pages} {113601} (\bibinfo {year} {2001})}\BibitemShut {NoStop}%
\bibitem [{\citenamefont {Ernst}\ \emph {et~al.}(2005)\citenamefont {Ernst}, \citenamefont {Samoson},\ and\ \citenamefont {Meier}}]{Ernst2005}%
  \BibitemOpen
  \bibfield  {author} {\bibinfo {author} {\bibfnamefont {M.}~\bibnamefont {Ernst}}, \bibinfo {author} {\bibfnamefont {A.}~\bibnamefont {Samoson}},\ and\ \bibinfo {author} {\bibfnamefont {B.~H.}\ \bibnamefont {Meier}},\ }\href {https://doi.org/10.1063/1.1944291} {\bibfield  {journal} {\bibinfo  {journal} {The Journal of Chemical Physics}\ }\textbf {\bibinfo {volume} {123}},\ \bibinfo {pages} {064102} (\bibinfo {year} {2005})}\BibitemShut {NoStop}%
\bibitem [{\citenamefont {Wang}\ \emph {et~al.}(2022)\citenamefont {Wang}, \citenamefont {Liu}, \citenamefont {Schloss}, \citenamefont {Alsid}, \citenamefont {Braje},\ and\ \citenamefont {Cappellaro}}]{Wang2022}%
  \BibitemOpen
  \bibfield  {author} {\bibinfo {author} {\bibfnamefont {G.}~\bibnamefont {Wang}}, \bibinfo {author} {\bibfnamefont {Y.-X.}\ \bibnamefont {Liu}}, \bibinfo {author} {\bibfnamefont {J.~M.}\ \bibnamefont {Schloss}}, \bibinfo {author} {\bibfnamefont {S.~T.}\ \bibnamefont {Alsid}}, \bibinfo {author} {\bibfnamefont {D.~A.}\ \bibnamefont {Braje}},\ and\ \bibinfo {author} {\bibfnamefont {P.}~\bibnamefont {Cappellaro}},\ }\href {https://doi.org/10.1103/PhysRevX.12.021061} {\bibfield  {journal} {\bibinfo  {journal} {Phys. Rev. X}\ }\textbf {\bibinfo {volume} {12}},\ \bibinfo {pages} {021061} (\bibinfo {year} {2022})}\BibitemShut {NoStop}%
\bibitem [{\citenamefont {Zarantonello}\ \emph {et~al.}(2019)\citenamefont {Zarantonello}, \citenamefont {Hahn}, \citenamefont {Morgner}, \citenamefont {Schulte}, \citenamefont {Bautista-Salvador}, \citenamefont {Werner}, \citenamefont {Hammerer},\ and\ \citenamefont {Ospelkaus}}]{Zarantonello2019}%
  \BibitemOpen
  \bibfield  {author} {\bibinfo {author} {\bibfnamefont {G.}~\bibnamefont {Zarantonello}}, \bibinfo {author} {\bibfnamefont {H.}~\bibnamefont {Hahn}}, \bibinfo {author} {\bibfnamefont {J.}~\bibnamefont {Morgner}}, \bibinfo {author} {\bibfnamefont {M.}~\bibnamefont {Schulte}}, \bibinfo {author} {\bibfnamefont {A.}~\bibnamefont {Bautista-Salvador}}, \bibinfo {author} {\bibfnamefont {R.~F.}\ \bibnamefont {Werner}}, \bibinfo {author} {\bibfnamefont {K.}~\bibnamefont {Hammerer}},\ and\ \bibinfo {author} {\bibfnamefont {C.}~\bibnamefont {Ospelkaus}},\ }\href {https://doi.org/10.1103/PhysRevLett.123.260503} {\bibfield  {journal} {\bibinfo  {journal} {Phys. Rev. Lett.}\ }\textbf {\bibinfo {volume} {123}},\ \bibinfo {pages} {260503} (\bibinfo {year} {2019})}\BibitemShut {NoStop}%
\bibitem [{\citenamefont {Deng}\ \emph {et~al.}(2024)\citenamefont {Deng}, \citenamefont {Li}, \citenamefont {Chen}, \citenamefont {Ni}, \citenamefont {Cai}, \citenamefont {Mai}, \citenamefont {Zhang}, \citenamefont {Zheng}, \citenamefont {Yu}, \citenamefont {Zou}, \citenamefont {Liu}, \citenamefont {Yan}, \citenamefont {Xu},\ and\ \citenamefont {Yu}}]{deng2024}%
  \BibitemOpen
  \bibfield  {author} {\bibinfo {author} {\bibfnamefont {X.}~\bibnamefont {Deng}}, \bibinfo {author} {\bibfnamefont {S.}~\bibnamefont {Li}}, \bibinfo {author} {\bibfnamefont {Z.-J.}\ \bibnamefont {Chen}}, \bibinfo {author} {\bibfnamefont {Z.}~\bibnamefont {Ni}}, \bibinfo {author} {\bibfnamefont {Y.}~\bibnamefont {Cai}}, \bibinfo {author} {\bibfnamefont {J.}~\bibnamefont {Mai}}, \bibinfo {author} {\bibfnamefont {L.}~\bibnamefont {Zhang}}, \bibinfo {author} {\bibfnamefont {P.}~\bibnamefont {Zheng}}, \bibinfo {author} {\bibfnamefont {H.}~\bibnamefont {Yu}}, \bibinfo {author} {\bibfnamefont {C.-L.}\ \bibnamefont {Zou}}, \bibinfo {author} {\bibfnamefont {S.}~\bibnamefont {Liu}}, \bibinfo {author} {\bibfnamefont {F.}~\bibnamefont {Yan}}, \bibinfo {author} {\bibfnamefont {Y.}~\bibnamefont {Xu}},\ and\ \bibinfo {author} {\bibfnamefont {D.}~\bibnamefont {Yu}},\ }\href {https://doi.org/10.1038/s41567-024-02619-5} {\bibfield  {journal} {\bibinfo  {journal} {Nature Physics}\ }\textbf {\bibinfo {volume} {20}},\ \bibinfo
  {pages} {1874} (\bibinfo {year} {2024})}\BibitemShut {NoStop}%
\bibitem [{\citenamefont {Deffner}\ and\ \citenamefont {Campbell}(2017)}]{Deffner2017}%
  \BibitemOpen
  \bibfield  {author} {\bibinfo {author} {\bibfnamefont {S.}~\bibnamefont {Deffner}}\ and\ \bibinfo {author} {\bibfnamefont {S.}~\bibnamefont {Campbell}},\ }\href {https://doi.org/10.1088/1751-8121/aa86c6} {\bibfield  {journal} {\bibinfo  {journal} {Journal of Physics A: Mathematical and Theoretical}\ }\textbf {\bibinfo {volume} {50}},\ \bibinfo {pages} {453001} (\bibinfo {year} {2017})}\BibitemShut {NoStop}%
\bibitem [{\citenamefont {Turchette}\ \emph {et~al.}(2000)\citenamefont {Turchette}, \citenamefont {Myatt}, \citenamefont {King}, \citenamefont {Sackett}, \citenamefont {Kielpinski}, \citenamefont {Itano}, \citenamefont {Monroe},\ and\ \citenamefont {Wineland}}]{Turchette2000}%
  \BibitemOpen
  \bibfield  {author} {\bibinfo {author} {\bibfnamefont {Q.~A.}\ \bibnamefont {Turchette}}, \bibinfo {author} {\bibfnamefont {C.~J.}\ \bibnamefont {Myatt}}, \bibinfo {author} {\bibfnamefont {B.~E.}\ \bibnamefont {King}}, \bibinfo {author} {\bibfnamefont {C.~A.}\ \bibnamefont {Sackett}}, \bibinfo {author} {\bibfnamefont {D.}~\bibnamefont {Kielpinski}}, \bibinfo {author} {\bibfnamefont {W.~M.}\ \bibnamefont {Itano}}, \bibinfo {author} {\bibfnamefont {C.}~\bibnamefont {Monroe}},\ and\ \bibinfo {author} {\bibfnamefont {D.~J.}\ \bibnamefont {Wineland}},\ }\href {https://doi.org/10.1103/PhysRevA.62.053807} {\bibfield  {journal} {\bibinfo  {journal} {Phys. Rev. A}\ }\textbf {\bibinfo {volume} {62}},\ \bibinfo {pages} {053807} (\bibinfo {year} {2000})}\BibitemShut {NoStop}%
\bibitem [{\citenamefont {Myatt}\ \emph {et~al.}(2000)\citenamefont {Myatt}, \citenamefont {King}, \citenamefont {Turchette}, \citenamefont {Sackett}, \citenamefont {Kielpinski}, \citenamefont {Itano}, \citenamefont {Monroe},\ and\ \citenamefont {Wineland}}]{Myatt2000}%
  \BibitemOpen
  \bibfield  {author} {\bibinfo {author} {\bibfnamefont {C.~J.}\ \bibnamefont {Myatt}}, \bibinfo {author} {\bibfnamefont {B.~E.}\ \bibnamefont {King}}, \bibinfo {author} {\bibfnamefont {Q.~A.}\ \bibnamefont {Turchette}}, \bibinfo {author} {\bibfnamefont {C.~A.}\ \bibnamefont {Sackett}}, \bibinfo {author} {\bibfnamefont {D.}~\bibnamefont {Kielpinski}}, \bibinfo {author} {\bibfnamefont {W.~M.}\ \bibnamefont {Itano}}, \bibinfo {author} {\bibfnamefont {C.}~\bibnamefont {Monroe}},\ and\ \bibinfo {author} {\bibfnamefont {D.~J.}\ \bibnamefont {Wineland}},\ }\href {https://doi.org/10.1038/35002001} {\bibfield  {journal} {\bibinfo  {journal} {Nature}\ }\textbf {\bibinfo {volume} {403}},\ \bibinfo {pages} {269} (\bibinfo {year} {2000})}\BibitemShut {NoStop}%
\bibitem [{\citenamefont {Bruzewicz}\ \emph {et~al.}(2019)\citenamefont {Bruzewicz}, \citenamefont {Chiaverini}, \citenamefont {McConnell},\ and\ \citenamefont {Sage}}]{Bruzewicz2019}%
  \BibitemOpen
  \bibfield  {author} {\bibinfo {author} {\bibfnamefont {C.~D.}\ \bibnamefont {Bruzewicz}}, \bibinfo {author} {\bibfnamefont {J.}~\bibnamefont {Chiaverini}}, \bibinfo {author} {\bibfnamefont {R.}~\bibnamefont {McConnell}},\ and\ \bibinfo {author} {\bibfnamefont {J.~M.}\ \bibnamefont {Sage}},\ }\href {https://doi.org/10.1063/1.5088164} {\bibfield  {journal} {\bibinfo  {journal} {Applied Physics Reviews}\ }\textbf {\bibinfo {volume} {6}},\ \bibinfo {pages} {021314} (\bibinfo {year} {2019})}\BibitemShut {NoStop}%
\bibitem [{\citenamefont {Michael}\ \emph {et~al.}(2016)\citenamefont {Michael}, \citenamefont {Silveri}, \citenamefont {Brierley}, \citenamefont {Albert}, \citenamefont {Salmilehto}, \citenamefont {Jiang},\ and\ \citenamefont {Girvin}}]{Michael2016}%
  \BibitemOpen
  \bibfield  {author} {\bibinfo {author} {\bibfnamefont {M.~H.}\ \bibnamefont {Michael}}, \bibinfo {author} {\bibfnamefont {M.}~\bibnamefont {Silveri}}, \bibinfo {author} {\bibfnamefont {R.~T.}\ \bibnamefont {Brierley}}, \bibinfo {author} {\bibfnamefont {V.~V.}\ \bibnamefont {Albert}}, \bibinfo {author} {\bibfnamefont {J.}~\bibnamefont {Salmilehto}}, \bibinfo {author} {\bibfnamefont {L.}~\bibnamefont {Jiang}},\ and\ \bibinfo {author} {\bibfnamefont {S.~M.}\ \bibnamefont {Girvin}},\ }\href {https://doi.org/10.1103/PhysRevX.6.031006} {\bibfield  {journal} {\bibinfo  {journal} {Phys. Rev. X}\ }\textbf {\bibinfo {volume} {6}},\ \bibinfo {pages} {031006} (\bibinfo {year} {2016})}\BibitemShut {NoStop}%
\bibitem [{\citenamefont {Wolf}\ and\ \citenamefont {Schmidt}(2021)}]{Wolf2021}%
  \BibitemOpen
  \bibfield  {author} {\bibinfo {author} {\bibfnamefont {F.}~\bibnamefont {Wolf}}\ and\ \bibinfo {author} {\bibfnamefont {P.}~\bibnamefont {Schmidt}},\ }\href@noop {} {\bibfield  {journal} {\bibinfo  {journal} {Measurement: Sensors}\ }\textbf {\bibinfo {volume} {18}},\ \bibinfo {pages} {100271} (\bibinfo {year} {2021})}\BibitemShut {NoStop}%
\bibitem [{Not()}]{Note1}%
  \BibitemOpen
  \href@noop {} {}\bibinfo {note} {H. Wu, C. Ho, G. Mitts, J. Rabinowitz and E. Hudson,  \url{https://github.com/EGGS-Experiment/Data/tree/main/Nonlinear-enhanced\%20wideband\%20sensing} (2026).}\BibitemShut {Stop}%
\end{thebibliography}
\end{document}